\documentclass[preprint,12pt]{elsarticle}

\usepackage{natbib}
\usepackage{lscape}
\usepackage{graphics}
\usepackage{times}
\usepackage{color}
\usepackage{colortbl}
 \usepackage{setspace}
\usepackage{caption}
\usepackage{epsfig,latexsym,graphicx}
\usepackage{psfrag}
\usepackage{comment}
\usepackage{amsthm}
\usepackage{amsmath,amscd,amssymb,amsfonts}
\usepackage{xspace}
\usepackage{listings}
\usepackage[justification=centering]{caption}
\usepackage{subfig}
\usepackage{mathtools}
\usepackage{bbm}
\usepackage{boxedminipage}
\usepackage{url}
\usepackage{cite}
\usepackage[ruled,vlined,linesnumbered]{algorithm2e}
\usepackage{multirow}

\journal{Elsevier}

\begin{document}

\begin{frontmatter}


 \title{Time varying channel estimation for RIS assisted network with outdated CSI: Looking beyond coherence time}
 \author[label1]{Souvik Deb \corref{cor1}}
\ead{deb.souvik5@gmail.com}
 \author[label1]{Sasthi C. Ghosh}
 \ead{sasthi@isical.ac.in}
\affiliation[label1]{organization={Advanced Computing \& Microelectronics Unit},
            addressline={Indian Statistical Institute},
             city={Kolkata},
            postcode={700108},
            state={West Bengal},
            country={India}}

\cortext[cor1]{Corresponding author}







\begin{abstract}
The channel estimation (CE) overhead for unstructured multipath-rich channels increases linearly with the number of reflective elements of reconfigurable intelligent surface (RIS). This results in a significant portion of the channel coherence time being spent on CE, reducing data communication time. Furthermore, due to the mobility of the user equipment (UE) and the time consumed during CE, the estimated channel state information (CSI) may become outdated during actual data communication. In recent studies, the timing for CE has been primarily determined based on the coherence time interval, which is dependent on the velocity of the UE. However, the effect of the current channel condition and pathloss of the UEs can also be utilized to control the duration between successive CE to reduce the overhead while still maintaining the quality of service. Furthermore, for muti-user systems, the appropriate coherence time intervals of different users may be different depending on their velocities. Therefore CE carried out ignoring the difference in coherence time of different UEs may result in the estimated CSI being detrimentally outdated for some users. In contrast, others may not have sufficient time for data communication. To this end, based on the throughput analysis on outdated CSI, an algorithm has been designed to dynamically predict the next time instant for CE after the current CSI acquisition. In the first step, optimal RIS phase shifts to maximise channel gain is computed. Based on this and the amount of degradation of SINR due to outdated CSI, transmit powers are allocated for the UEs and finally the next time instant for CE is predicted such that the aggregated throughput is maximized. Simulation results confirm that our proposed algorithm outperforms the coherence time-based strategies.
\end{abstract}
\begin{keyword}

Reconfigurable intelligent surface \sep channel coherence time\sep outdated channel state information \sep power allocation \sep throughput optimization \sep 5G.



\end{keyword}

\end{frontmatter}

\section{Introduction}
Reconfigurable intelligent surface (RIS) is a promising solution to increase coverage and spectral efficiency for enhanced mobile broadband services (eMBB) in RIS-assisted 5G cellular systems \citep{RIS_model,WPC_RIS_NOMA_select,modelling_RIS,RIS_6G}. An RIS comprises a uniform rectangular array (URA) of meta-materials called reflective elements. These reflecting elements control the phase shifts of the incident electromagnetic wave such that the reflected signal is added constructively or destructively with other received signals to enhance the spectral efficiency of the user equipment (UE). However, to optimally use an RIS for improving spectral efficiency \citep{D2D_ris}, \citep{ris_mmwave}, \citep{WPC_RIS_NOMA_select} and coverage \citep{ris_vehicular}, \citep{irs}, perfect channel state information (CSI) is assumed at the new radio gNodeB (gNB). However, channel estimation (CE) for unstructured channels consumes a significant amount of time since the pilot overhead in RIS-assisted networks for multi-path rich channels increases linearly with the number of reflecting elements \citep{CE_framework}. Furthermore, UEs are mobile in the network environment. As a consequence, the estimated channel becomes outdated and perfect CSI is no longer available.     

The gNB in an RIS-assisted network communicates data during the time between consecutive CE phases. The RIS itself is incapable of CE for obtaining the CSI. The gNB is responsible for CE through feedback links resulting in large delays \citep{outdated_infocom}. RISs with a large number of elements consume a significant portion of the time allotted for data transmission to send pilot signals required for accurate CE \citep{fast_ce_ris}, \citep{ce_practicle_ris}, \citep{ce_ris_mimo}. This results in only a fraction of the allotted time for data communication, reducing the achievable throughput. Obtaining perfect instantaneous CSI for the case of high mobile UEs entails massive signalling overhead and delays. For high mobility UE, the rapid changes in channel conditions cause the estimated CSI to become outdated due to feedback delays \citep{RIS_outdate_snr}. This makes it essential to consider the performance of an RIS-assisted 5G network with outdated CSI as opposed to assuming that perfect instantaneous CSI is always available. 

\begin{figure}
    \centering
    \includegraphics[scale=0.6]{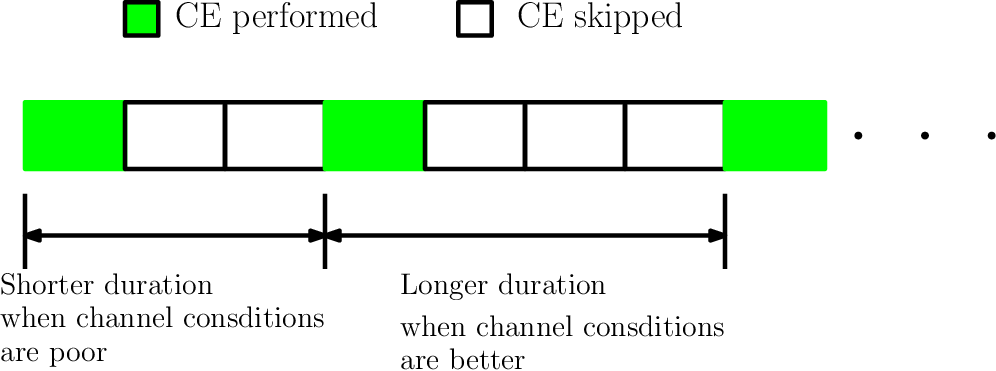}
    \caption{Time varying channel estimation}
    \label{tvce}
\end{figure}

Keeping outdated CSI in mind, one of the most essential questions is when to perform CE? The CE timing in recent studies is based only on the channel coherence time. The estimated channel conditions are assumed to be constant throughout the coherence time and the channel is estimated again once the coherence time runs out. However, there is still a non-zero delay between the estimated channel and the actual channel during data communication even within the coherence time interval. Moreover, the computation of channel coherence time is dependent on the velocity of the UE. A mobile UE changes its position with respect to the serving gNB and the neighbouring cells over time. The velocity of the UE can also change over time. Consequently, the pathloss, fading and background interference experienced by a UE varies over time. Furthermore, different UEs in a multi-user system move at different velocities thereby leading to different coherence time intervals experienced by each UE. Therefore, the rate of degradation in datarate due to delay between outdated CSI and actual CSI varies differently for each UE over time. Moreover, the power allocated for each UE also determines the amount of degradation of signal to interference plus noise ratio (SINR) suffered by them. Therefore intelligent transmit power sharing among the UEs can improve the aggregate throughput \citep{OFDMA_embb_urllc}. This makes it essential to capture the effects of channel condition, interference, phase shift values of the RIS and power allocation along with velocity when determining the timing for CE as described in Fig. \ref{tvce}. Considering that the time is discritized into slots, CE estimation can be skipped in certain consecutive slots depending on the channel conditions. Fig. \ref{tvce} shows that by skipping the required number of time slots the duration between consecutive CE phases can be dynamically adjusted. The shaded time slots are the slots when CSI is acquired. Based on the quality of the channel, the CE phase is skipped in certain number of the consecutive slots marked as white slots.

There have been several recent studies on minimizing the CE overhead. In \citep{optimal_group_strategy} and \citep{ris_grouping_q_learning}, the authors have proposed an optimal RIS elements grouping to enhance throughput by reducing pilot overhead. In \citep{performance_grouping}, the authors have analysed the effect of RIS elements grouping on CE overhead. An analytical expression has been derived in \citep{LIS_optimal_on/off} to find the optimal number of RIS elements to switch on in a large intelligent surface (LIS) for reducing pilot overhead. In \citep{csi_discrete_phase_shift}, a hierarchical training reflection design has been proposed to estimate the RIS channel utilizing RIS-element grouping and partition. However, even with reduced pilot overhead the estimated channel still becomes outdated as the UE continuously moves. The studies cited above do not consider the fact that due to the mobility of the UE the estimated channel may become outdated.  

Recently many studies have also been done to understand the impact of outdated CSI on RIS-assisted network performance. In \citep{RIS_outdate_snr}, the authors propose an RIS deployment mechanism in centralized and distributed cases for outdated CSI. In \citep{outdated_infocom}, authors derive a closed form expression for the effective capacity for RIS-enabled millimeter wave (mmWave) downlink. In \citep{UAV_RIS_selection}, an RIS selection strategy has been proposed for unmanned aerial vehicle (UAV) based multi-user downlink network with imperfect and outdated CSI. In \citep{RIS_outdate_secrecyrate}, authors enhance the secrecy rate by jointly optimizing the transmit power and RIS's reflective beamforming considering outdated CSI. The above-cited studies do not provide any insight on the CE timing for the scenario of outdated CSI. Furthermore, all the above cited studies only consider a fixed channel coherence time depending only on the velocity of a UE \citep{lower_bound_coherence_time} as the interval between two consecutive CE.

In \citep{time_varying_CSI}, authors consider time-varying cascaded CE over RIS-assisted communication using structured channel model. They use deep learning for channel extrapolation in both time and antenna domains. However, they do not provide any particular strategy for computing the exact timing for CE. Moreover, the effect of channel conditions, interference and power allocation have not been captured for determining CE timing. \emph{To the best of our knowledge, no recent studies have addressed time-varying channel estimation for RIS-assisted networks  under the conditions of unstructured channels and outdated CSI.} To this end, the major contribution of this work is to develop an algorithm that takes into account the channel conditions and velocities of the UEs during the last channel estimation phase and  determines phase shifts of RIS, power allocation for each UE and accordingly predict the next time instant for CE such that the aggregated throughput is maximized. Specifically, our contributions in this work are summarized as follows:

\begin{itemize}

    \item Firstly, for each time slot when CE is not performed, we optimize the phase shift values of the RIS elements based on the outdated channel to maximize the channel gain at each UE using semi-definite programming. 
    
    \item  
    After the phase shift values for the RIS elements are computed at a given time slot, we propose an algorithm to intelligently allocate transmit powers to the UEs based on the outdated channel to maximise the aggregate throughput.

     \item Finally, we propose an algorithm to dynamically adjust the time interval between consecutive CE that utilises the tradeoff between tolerating degradation due to delay of outdated CSI and saving time spent on CE. It then predicts the number of consecutive slots during which CE can be skipped to maximize the aggregate throughput gain. Here throughput gain achieved by a UE is defined as the difference between the actual throughput achieved by the UE without performing any CE and the expected throughput of the UE if CE had been performed. 
    
    \item We perform extensive system level simulations and show that the proposed algorithm outperforms the existing coherence time based channel estimation approaches in terms of aggregated throughput. We also study the distribution of the number of time slots skipped between two consecutive CE using box plots.

\end{itemize}

The rest of the manuscript is organized as follows. The considered system model has been described in section \ref{sysmodel}. In section \ref{pf}, we describe our proposed time varying CE estimation algorithm. In this section, we define the aggregate throughput gain, optimise RIS phase shifts, determine transmit powers for each UE and finally propose an algorithm to predict the number of time slots for which CE need not be performed. In section \ref{sim}, simulation results have been presented. Finally, section \ref{con} concludes the paper. Important symbols used in the manuscript are summarized in Table \ref{symtab}. 

\section{System model}\label{sysmodel}

\subsection{Environment model}

We consider a downlink wireless network with a single gNB providing ubiquitous coverage to a set $\mathcal{U}=\{1,\;2,\;\cdots\;U\}$ of eMBB UEs. The total bandwidth is divided into $B$ resource blocks (RBs). The UEs adopt orthogonal frequency division multiple access (OFDMA) and each UE has $b=\frac{B}{U}$ RBs \citep{OFDMA_embb_urllc}. The total allowed transmit power at the gNB is $P_{tot}$. Moreover, the gNB communicates with the UEs over discrete time slots. The UEs receive interfering signals from neighbouring cells. The gNB communicates with the UEs utilizing the assistance of an RIS with $M$ passive reflective elements.  The gNB is at a distance $D$ from the centre of the RIS. The service range of the RIS is considered to be $R$ meters.  We consider a 2-D scenario where the UEs roam on the ground level plane with the same antenna height at all positions. The coherence time for UE $u$ denoted by $T_{uc}$ is defined as the maximum time interval over which the time correlation of the channel is above a predefined threshold $\rho^{th}$ \citep{rappaport2024wireless} \citep{lower_bound_coherence_time}. Let $f_{ud}=f_cV_u/c$ is the maximum Doppler spread, $f_c$ is the carrier frequency, $c$ is the velocity of light and $V_u$ is the velocity of the UE $u$ then $T_{uc}$ is approximately computed as follows \citep{lower_bound_coherence_time}:
\begin{equation}\label{timesloteq}
    T_{uc}=\frac{c\cos^{-1}(\rho^{th})}{2\pi V_uf_{c}}.
\end{equation}
 We consider that time is discritized into time slots of length $T_c=\displaystyle\min_{u\in\mathcal{U}}\{T_{uc}\}$. The corresponding coherence block length is computed as $N_c=T_cB_c$, where $B_c=\frac{\cos^{-1}(f_c)}{2\pi\tau_{max}}$ is the coherence bandwidth for a maximum delay spread of $\tau_{max}$.


\begin{figure}[t]
    \centering
    \includegraphics[scale=0.6]{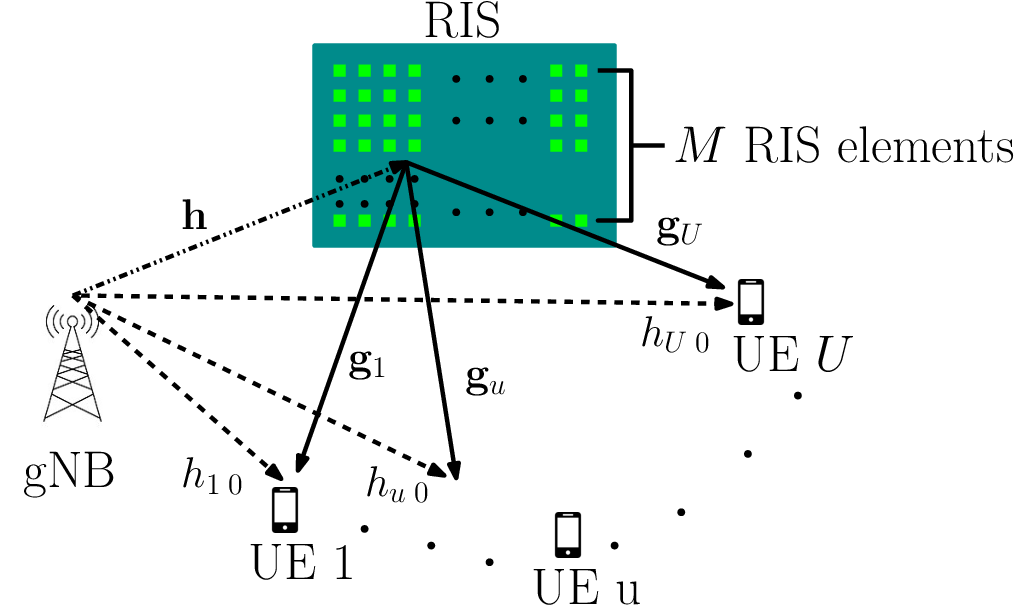}
    \caption{System model}
    \label{fsysmod}
\end{figure}

\subsection{Outdated channel and SINR model}

\begin{table}[h!]
\footnotesize
    \centering
    \caption{Symbol Table}
    \label{symtab}
    \begin{tabular}{|c|c|c|c|}
    \hline
       \textbf{Symbol}  & \textbf{Meaning} &\textbf{Symbol}  & \textbf{Meaning}\\\hline
        $U$  & No. of UEs& $\mathbf{\hat{g}}_u$ & \begin{tabular}{@{}c@{}}Outdated channel \\ from RIS to UE $u$\end{tabular} \\\hline
        $\hat{h}_{u0}$ &  \begin{tabular}{@{}c@{}}Outdated channel \\ from gNB to UE $u$\end{tabular} &$\lambda_c$ & carrier wavelength\\\hline
        $\mathcal{U}$& Set of UEs& $T_{d0}$& \begin{tabular}{@{}c@{}}Estimation delay\\ of direct channel\end{tabular}\\\hline
        $B$ & No. of resource blocks&$T_d$ &\begin{tabular}{@{}c@{}}Estimation delay\\ of cascaded channel\end{tabular}\\\hline
        $W$ & Bandwidth of each UE& $\rho_{u0}$&\begin{tabular}{@{}c@{}}Correlation coefficient\\ between outdated and\\ actual direct channel\end{tabular}\\\hline
        $\hat{h}_{u0}$ & \begin{tabular}{@{}c@{}}Outdated direct channel\\ from gNB to UE $u$\end{tabular} &$\rho_u$&\begin{tabular}{@{}c@{}}Correlation coefficient \\between outdated and\\ actual cascaded channel\end{tabular}\\\hline
        $D$ & BS-RIS center distance&$V_u$ & velocity of UE $u$\\\hline
        $R$ & Range of RIS service&$f_c$ & Carrier frequency\\\hline
        $M$ & No. of RIS elements&$f_{ud}$ &Doppler spread for UE $u$\\\hline
        $\Psi$ &\begin{tabular}{@{}c@{}} Coherence time based\\interval between \\ consecutive CE\end{tabular} &$c$& Speed of light\\\hline
        
        $\mathbf{h}$ & Channel from gNB to RIS& \begin{tabular}{@{}c@{}}$\Omega_0,\;\Omega_1,$\\$\;\Omega_2$\end{tabular}&\begin{tabular}{@{}c@{}} Mean of Rician variables\\ $|\hat{h^u}_0|$, $|h_m|$ and $|\hat{g_u}_m|$\end{tabular}\\\hline
        $\mathbf{g}_u$ & Channel from RIS to UE $u$ &$\Delta_u$& Expected SINR at UE $u$\\\hline
        
        $h_{u0}$ & Channel from gNB to UE $u$&$G_n$ & \begin{tabular}{@{}c@{}} Aggregate throughput gain \\ at time slot $l+n$ \end{tabular}\\\hline
        $\mathbf{\Phi}$ & Phase shift matrix of RIS& $\mathbf{L}$ & Pathloss matrix : gNB-RIS-UE\\\hline
        
       $\Sigma_u$& \begin{tabular}{@{}c@{}}Standard deviation \\of interference\end{tabular}&$N_{max}$&\begin{tabular}{@{}c@{}} Smallest integer at which\\channel correlation becomes 0\end{tabular}\\\hline

        $\sigma_{uf}$ & Outdated CSI noise at UE $u$ & $\tau_{max}$ & Maximum delay spread\\\hline

        $R(l)$ & Throughput at time slot $l$&$\bar R(l)$& Expected throughput at slot $l$\\\hline

        $\beta_0$& Pathloss : gNB-UE& $\rho^{th}$ &Time correlation threshold\\\hline
        $\beta_u$ & Pathloss : gNB-RIS-UE $u$ &$T_c$ & Time slot duration\\\hline
        
        $n_{u0}$ & Additive white Gaussian noise& $\mathbf{\bar P}$ &\begin{tabular}{@{}c@{}}Transmit power vector\\ for expected throughput\end{tabular}\\\hline
        
        $\mathbf{I}_u$& \begin{tabular}{@{}c@{}}Interference from \\neighbouring cells\end{tabular}& $T_p$ & Duration of pilot overhead\\\hline
        $\delta_u$ & Actual SINR at UE $u$ & $\mathbf{P}$& Transmit power vector\\\hline
        $G_u$ & Gain of throughput for UE $u$&$P_{tot}$& Total transmit power allowed\\\hline

    \end{tabular}

\end{table}

We consider that the UEs lie in the far-field of the RIS. The phase shift induced by the reflective elements of the RIS is represented through a diagonal matrix as follows \citep{ris_mmwave}:
\begin{equation}
\mathbf{\Phi}=diag(e^{j\phi_1},\;e^{j\phi_2},\; \dots ,\;e^{j\phi_{M}}),
\end{equation}
\noindent where, $\phi_m \in [0,2\pi]$ for all $m=1,\;2,\;\cdots,\;M$. Let $x_u(t)$ be the transmitted signal from the gNB to the UE $u$ at time $t$. Let $h_{u0}\in \mathbb{C}^1$, $\mathbf{h}=[h_1,\;h_2,\cdot\cdot,h_m\cdots,h_M]^T$ $\in \mathbb{C}^{M}$ and $\mathbf{g}_u=[g_{u1},\;g_{u2},\cdot\cdot,g_{um},\cdots,\;g_{uM}]\in\mathbb{C}^{1\times M}$ be the channel vectors from gNB to UE $u$,  gNB to RIS and from RIS to UE $u$ respectively, following Rician fading models \citep{RIS_outdate_snr}. A schematic of the system model is presented in Fig. \ref{fsysmod}. Due to the mobility of UEs and the delay caused by the massive number of pilot transmissions during channel estimation \citep{CSI_overhead} \citep{CSI_2021_large}, it is considered that the CSI for the direct gNB-UE channel and RIS-UE  channel is outdated with $T_{d0}$ and $T_d$ being the respective estimation delays. The relation between the actual channels $h_{u0}$ and $\mathbf{g}_u$ and the outdated channels $\hat{h}_{u0}$ and $\mathbf{\hat{g}}_u=[\hat{g}_{u1},\;\hat{g}_{u2},\;\cdots,\hat{g}_{um},\cdots,\;\hat{g}_{uM}]\in\mathbb{C}^{1\times M}$ is as follows \citep{RIS_outdate_snr}: 
\begin{align}
& h_{u0}=\rho_{u0} \hat{h}_{u0}+\bar{\rho}_{u0} \omega_0,\label{direct_eq} \\
& \mathbf{g}_u=\rho_u \mathbf{\hat{g}}_u+\bar{\rho_u} \boldsymbol{\omega}.\label{ref_eq}
\end{align}
\noindent Here, $0\le\rho_{u0}\le 1$ and $0\le\rho_u\le 1$ are the correlation coefficients between the outdated channel estimate and the actual channel given by $\rho_{u0}=J_0\left(2 \pi f_{ud}, T_{d0}\right)$ and $\rho_u=J_0\left(2 \pi f_{ud}, T_d\right)$ \citep{RIS_outdate_snr},
where $J_0(.)$ is the zeroth order Bessel's function of the first kind. Moreover, $\boldsymbol{\omega}=[\omega_{u1},\;\omega_{u2},\;\cdots,\;\omega_{uM}]\in\mathcal{C}^{1\times M}$ where $\omega_{ui}\sim\mathcal{CN}(0,\sigma^2_{\hat{g}_{ui}})$ and $\omega_0\sim \mathcal{CN}(0,\sigma^2_{\hat{h}_{u0}})$. Furthermore, $\bar{\rho}_{u0}=\sqrt{1-\rho_{u0}^2}$ and $\bar{\rho_u}=\sqrt{1-\rho_u^2}$. Therefore the received signal for the outdated CSI is as follows \citep{RIS_outdate_snr}:
\begin{eqnarray}
y_u(t)=\sqrt{P_u}(\rho_u\hat{\mathbf{g}}_u\mathbf{L}_u\mathbf{\Phi}\mathbf{h}+\rho_{u0}\sqrt{\beta_{u0}^{-1}}\hat{ h}_{u0})x_u(t)\nonumber\\
+(\bar{\rho_u}\boldsymbol{\omega}\mathbf{L}_u\mathbf{\Phi}\mathbf{h}+\bar{\rho}_{u0}\sqrt{\beta_{u0}^{-1}}\omega_0)x_u(t)+n_{u0}+ \mathbf{I}_u,
\end{eqnarray}
\noindent where $n_{u0}\in\mathcal{N}(0,\sigma_u^2)$ is the additive white Gaussian noise (AWGN) with $0$ mean and variance $\sigma_u^2$ and $P_u$ is the power associated with $x_u(t)$ allocated to UE $u$. Adopting from \citep{tong2022two_stage}, $\mathbf{I}_u\sim \mathcal{LN}(\mu_u,\Sigma_u^2)$ is the interference suffered by UE $u$ from neighbouring cells and follows log normal distribution with mean $\mu_u$ and standard deviation $\Sigma_u$. Here $\beta_{u0}$ is the pathloss of the direct channel from gNB to UE $u$ and the pathloss of the cascaded gNB-RIS-UE channel is represented using the diagonal matrix $\mathbf{L}_u=diag(\sqrt{\beta_{u}^{-1}},\;\sqrt{\beta_{u}^{-1}},\cdots,\;\sqrt{\beta_{u}^{-1}})\in \mathbb{R}^{M\times M}$ \citep{RIS_outdate_snr}.
 The mean of the Rician random variables $|\hat{h}_{u0}|$, $|h_m|$ and $|\hat{g}_{um}|$ are denoted by $\Omega_0$, $\Omega_1$ and $\Omega_2$ respectively. The effective noise (outdated CSI noise + AWGN noise) is computed as follows:
\begin{equation}
    \sigma_{uf}^2(\rho_{u0},\rho_u)=\frac{P_u M \bar{\rho_u}^2\left(1-\Omega_2^2\right)}{\beta_u}+\frac{P_u \bar{\rho_u}_0^2\left(1-\Omega_0^2\right)}{\beta_{u0}}+\sigma_u^2.
\end{equation}

\noindent Therefore, adopting the interference computation from \citep{tong2022two_stage}, $\mathbb{E}(\mathbf{I}_u)=\exp(2\mu_u+2\Sigma_u^2)$, and the received SINR at UE $u$ for the channel correlation coefficients $\rho_u$ and $\rho_{u0}$ is computed as follows:
\begin{equation}\label{SINR_actual}
\delta_u(\rho_{u0},\rho_u)=\frac{P_u|\rho_u\hat{\mathbf{g}}_u\mathbf{L}_u\mathbf{\Phi}\mathbf{h}+\rho_{u0}\sqrt{\beta_{u0}}^{-1}\hat{h}_{u0}|^2}{\sigma_{uf}^2+\exp(2\mu_u+2\Sigma_u^2)}.
\end{equation}
\noindent The reflection phase of the RIS can be optimized using algorithms presented in \citep{OFDMA_embb_urllc}.
To compute the expected SINR at any time $t$ for the channel correlation coefficients $\rho_u$ and $\rho_{u0}$ we adopt the value of phase difference in the received signal for optimized phase shift values of the RIS from \citep{UAV_RIS_selection} and the expected SINR defined as $\Delta_u$ is given as follows \citep{RIS_outdate_snr}:
\begin{align}\label{esinr}
  \Delta_u=&\Delta_{uf}\big(M(\rho_u)^2/\beta_u + M(M-1)(\rho_u\Omega_1\Omega_2)^2/\beta_u\nonumber\\
   &+2M\rho_u\rho_{u0}\Omega_0\Omega_1\Omega_2/\sqrt{\beta_{u0}\beta_u}+\rho_{u0}^2\beta_{u0}^{-1}\big),
\end{align}

\noindent where $\Delta_{uf}=\frac{P_u}{\sigma_{uf}^2(\rho_{u0},\rho_u)+\mathbb{E}(\mathbf{I}_u)}$.

For a given time slot $l$, we use \eqref{SINR_actual} to compute the actual throughput of a UE at time slot $l$. Had CE been performed at the beginning of the time slot $l$, we compute the expected throughput the UE would achieve assuming outdated CSI using \eqref{esinr}. 



\section{The proposed time varying CE}\label{pf} 
In this section, for a given time slot, we first compute the throughput gain achieved by a UE without CSI acquisition as opposed to the same after CSI acquisition. Then we compute the optimal phase shift of the RIS to maximize the channel gain at each UE and based on it we formulate the power allocation problem as a convex optimization problem which maximizes the aggregate throughput gain of all the UEs in the network. Finally, utilizing the gain computed for consecutive future time slots after the last slot in which CE was performed, an algorithm is proposed to compute the number of consecutive future time slots where CE can be skipped. 

\subsection{Throughput gain}

\begin{figure}
    \centering
    \includegraphics[scale=0.5]{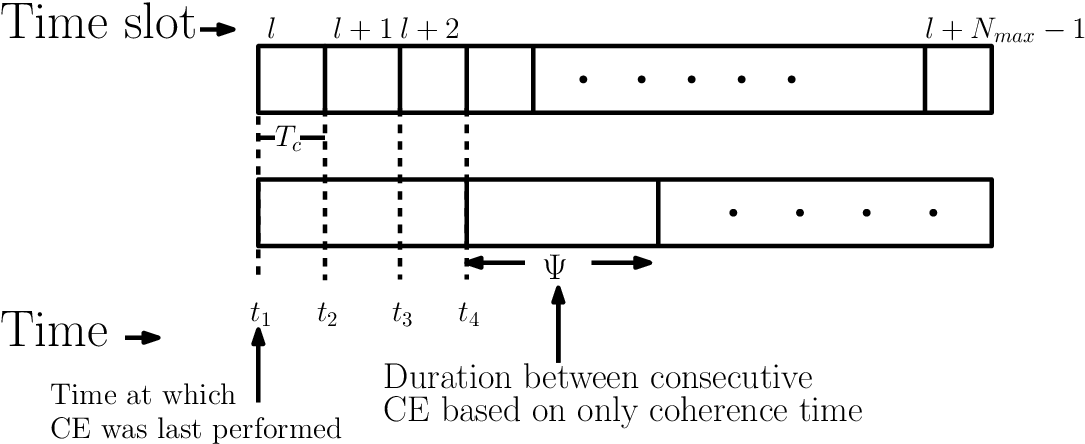}
    \caption{Time slot division}
    \label{time}
\end{figure}

 In this subsection, we define the gain achieved by a UE at time slot $l$ as the difference between the actual throughput without CSI acquisition and the expected throughput with CSI acquisition at the beginning of slot $l$.
Let $T_p\geq M+1$ \citep{optimal_group_strategy} denote the duration of pilot overhead during CSI acquisition of a UE. It is to be noted that the time left for actual data communication within a time slot is $(1-\frac{T_p}{N_c})\times T_c$. Moreover, due to the massive pilot overhead for large RISs and mobility of the UE the CSI acquired by the UE becomes outdated by a time delay of at least $\frac{T_p}{N_c}T_c$. 
Therefore, the correlation coefficient between the estimate of the channel and the actual channel during data communication at time slot $l$ for UE $u$ is denoted by $\rho_{u0}[l]$ for the direct gNB-UE channel and $\rho_u[l]$ for the cascaded gNB-RIS-UE channel. Let $l$ be the last time slot when CE was performed and $t_1$ be the time instant when time slot $l$ begins as shown in Fig. \ref{time}. We index this particular time slot as $l+0$. Furthermore, let $V=\displaystyle\max_{u\in\mathcal{U}}{V_u}$ be the maximum velocity achieved by a UE in the network and $f_d=f_cV/c$ be the corresponding maximum Doppler spread. Hence, $\rho_{u0}[l]=\rho_u[l]=J_0(2\pi f_d,t_1+\frac{T_p}{N_c}T_c)$. The $n^{th}$ time slot after $l$ is denoted by $l+n$. Therefore, the delay between the actual channel at time slot $l+n$ and the outdated channel estimated at time slot $l$ is $nT_c$ for all $n\geq1$. The time correlation of the channel corresponding to $f_d$ is computed as $\rho_u[l+n]=\rho_{u0}[l+n]=J_0(2\pi f_d,t_1+nT_c)$. It may be noted that when $\rho'_u=\rho'_{u0}=0$, it means that the time delay between the current channel and the estimated channel is long enough for them to become completely uncorrelated and hence CE has to be performed again to estimate the new channel which is completely independent of the previously estimated channel. We define $N_{max}=\displaystyle\min\{n\in\mathbb{Z}^+|\; J_0\left(2 \pi f^u_d, t_1+nT_c\right)\le0 \}$. This means that $l+N_{max}$ is the first time slot after $l$ where $\rho'_u=\rho'_{u0}=0$. Furthermore, since the time delay between $l$ and $l+N_{max}$ is in the order of milliseconds, we assume the location of the UE to be approximately the same during slot $l$. Finally, we denote the set of consecutive time slots after the slot $l$ upto the time slot $l+ N_{max}-1$ as $\mathcal{T}=\{1,\;2,\;3\;\cdots\;N_{max}-1\}$. For time slots after $l+N_{max}-1$, the SINR values computed in \eqref{SINR_actual} and \eqref{esinr} become too erroneous to be useful.

Now, from \eqref{SINR_actual} the  throughput for UE $u$ is computed using Shannons's capacity formula as follows:

\begin{equation}\label{rate_u}
R_u(l+n)=\left(1-\frac{\alpha_lT_p}{N_c}\right)W\log_2\left(1+\delta_u(\rho^u_0[l+n],\rho^u[l+n])\right).    
\end{equation}
\noindent Using \eqref{esinr}, the upper bound of the expected capacity of the UE $u$ at time slot $l$ can be computed as follows:
\begin{align}\label{erateu}
    \bar R_u(l+n)=\left(1-\frac{\bar\alpha_lT_p}{N_c}\right)W\log_2(1+\Delta^u(\rho^u_0[l+n],\rho^u[l+n]). 
\end{align}
\noindent Here, $W$ is the allocated bandwidth and $\alpha_l=\bar\alpha_l=0$ if CE is skipped at time slot $l$. Otherwise $\alpha_l=\bar\alpha_l=1$.
It is clear from \eqref{rate_u} and \eqref{erateu} that the throughput for each UE can vary based on the channel conditions (pathloss and background interference) and velocity even for the same amount of delay between the current channel and channel estimated during the last CE process. It may also be noted that skipping the CE phase for some time slots reduces the total time spent on CSI acquisition thereby increasing time for data communication. However, delaying the CE causes the SINR to degrade as the CSI becomes more outdated. 


Now we are prepared to define the throughput gain achieved by a UE $u$ at time slot $l+n$ where $1\leq n \leq N_{max}-1$.
In recent works, such as \citep{performance_grouping}, \citep{optimal_group_strategy}, \citep{RIS_6G}, \citep{RIS_UAV_ergodic_throughput_max}, \citep{UAV_RIS_selection} the time duration between two consecutive CE processes is determined solely based on coherence time and channel time correlation threshold 
$\bar \rho^{th}\leq \rho^{th}$ \citep{linear_RIS_CE}, \citep{lower_bound_coherence_time}. Let such a time duration be denoted by $\Psi $. Therefore, the number of time slots that approximately make up $\Psi$ is computed as 
$\lfloor{\frac{\Psi}{T_c}}\rfloor$.   
It may be noted that depending on the channel conditions and position of different UEs the outdated CSI may degrade the SINR to the extent that it becomes beneficial to perform CE even before $\Psi$. For example in Fig. \ref{time}, $\Psi$ is composed of $3$ time slots. Therefore, for the case of fixed CE timing based only on channel coherence time, the next CE phase after the one at $t_1$ happens at $t_4$ since $\Psi=t_4-t_1$. On the other hand, when time varying CE is allowed, CSI can be acquired either at $t_2$ or $t_3$ or even at a time beyond $t_4$ depending on the channel conditions of each UE measured at $t_1$. On the other hand, an intelligent allocation of transmit power for each UE can allow UEs to achieve greater throughput by utilizing the time saved by avoiding CE even after $\Psi$. In either case, taking the effect of channel conditions of different UEs into consideration the time for CE can be dynamically adjusted to increase the aggregate throughput gain. From \eqref{rate_u} and \eqref{erateu} we define the following:

\begin{equation}\label{gain}
    G_u = \begin{cases}
        \max\left(\bar{R}_u(l) - R_u(l+n), 0\right) & \text{if } 1 \leq n \leq \left\lfloor \frac{\Psi}{T_c} \right\rfloor \\[8pt]
        R_u(l+n) - \bar{R}_u\left(l+n \mod \left\lfloor \frac{\Psi}{T_c} \right\rfloor\right) & \text{if } \left\lfloor \frac{\Psi}{T_c} \right\rfloor + 1 \leq n \leq N_{\text{max}} - 1
    \end{cases}
\end{equation}

\noindent Furthermore, the values of $\alpha_l$ and $\bar \alpha_l$  for $0\leq n\leq N_{max}-1$ are given as follows:
\begin{equation}
    \bar \alpha_{l+n}=\begin{cases}
        1 & \text{If } n \mod  \left\lfloor \frac{\Psi}{T_c} \right\rfloor\ =0\\[8pt]
        0 & \text{Otherwise},
    \end{cases}
\end{equation}
\begin{equation}
    \alpha_{l+n}=\begin{cases}
        1 & \text{If } n=0\\[8pt]
        0 & \text{Otherwise}.
    \end{cases}
\end{equation}
\noindent Therefore the aggregate throughput gain for all UEs at time slot $l+n$ in the network is defined as 
\begin{equation}\label{sumgain}
    G_n=\displaystyle\sum_{u=1}^U G_u.
\end{equation}

\subsection{Phase shift optimization }

 The correlation coefficient between the channel estimated at time slot $l$ and the actual channel at slot $l+n$ for UE $u$ be $\rho^u_0[l+n]$ for the direct gNB-UE channel and $\rho^u[l+n]$ for the cascaded gNB-RIS-UE channel. We first determine the phase shift of the RIS elements in order to maximize the channel gain at each UE. We adopt the same optimization strategy presented in \citep{OFDMA_embb_urllc} to compute the optimal phase shifts. However, \citep{OFDMA_embb_urllc} does not consider outdated CSI. In order to compute the optimal phase shifts for outdated CSI, the channel gain at UE $u$ at time slot $l+n$ where $n\in \mathcal{T}$ in \eqref{SINR_actual} is given by 
 \begin{align*}
|\rho_u[l+n]\mathbf{\hat{g}}_u\mathbf{L}_u\mathbf{\Phi}\mathbf{h}+\rho_{u0}[l+n]\sqrt{\beta_{u0}^{-1}}h_{u0}|^2=|\rho_u[l+n]\hat{\mathbf{g'}}_u\mathbf{\Phi}\mathbf{h}\\
+ \rho_{u0}[l+n]\sqrt{\beta_{u0}^{-1}}h_{u0}|^2.
\end{align*}
Here $\hat{\mathbf{g'}}_u=[\sqrt{\beta_u^{-1}}\hat{g}_{u1},\;\sqrt{\beta_u^{-1}}\hat{g}_{u2},\;\cdots,\;\sqrt{\beta_u^{-1}}\hat{g}_{uM}]\in\mathbb{C}^{1\times M}$. Let $s_m=e^{\phi_m}$ and $\mathbf{s}=[s_1,\;s_2,\;\cdots,\;s_M]^T\in\mathbb{C}^{M}$. The channel gain can be rewritten as $$|\mathbf{s}^H\boldsymbol{\Theta}+\rho_{u0}[l+n]\sqrt{\beta_{u0}}^{-1}\hat{h}_{u0}|^2.$$ Here $\boldsymbol{\Theta}=diag(\rho_u[l+n]\hat{\mathbf{g'}}_u)\mathbf{h}\in\mathbb{C}^M$. 

 The first step is to find the optimal phase shift vector $\mathbf{s}$ such that the minimum channel gain among all UEs is maximised. This can be formulated as the optimization problem as follows:

 \begin{align}\label{phaseopt}
     &\displaystyle\max_{\mathbf{s}} \displaystyle\min_{u\in\mathcal{U}}|\mathbf{s}^H\boldsymbol{\Theta}+\rho_{u0}[l+n]\sqrt{\beta_{u0}}^{-1}\hat{h}_{u0}|^2\\
     & \text{s.t.  } |s_n|=1\;\forall 1\leq n\leq M.
 \end{align}

\noindent Consider the following matrix as defined in \citep{OFDMA_embb_urllc}:

\begin{equation}
    \mathbf{X}_u=\begin{bmatrix}
    \boldsymbol{\Theta}\boldsymbol{\Theta}^H & \boldsymbol{\Theta}\rho_{u0}[l+n]\sqrt{\beta_{u0}^{-1}}\hat{h}^{\oplus}_{u0}\\
    \rho_{u0}[l+n]\sqrt{\beta_{u0}^{-1}}\hat{h}_{u0}\boldsymbol{\Theta}^H & 0
    \end{bmatrix}.
\end{equation}
\noindent where $(.)^{\oplus}$ is the complex conjugate. 
Let $\mathbf{\bar s}=[s_1,\;s_2,\;\cdots,\;s_M,x]^T$, where $x$ is an auxiliary variable. Defining  $\mathbf{S}=\mathbf{\bar s}\mathbf{\bar s}^H$, adding an auxiliary variable $\xi$ and  adopting the transformation in \citep{OFDMA_embb_urllc}, we can rewrite the optimization formulation in \eqref{phaseopt} as a semi-definite programming (SDP) as follows:

\begin{align}\label{SDR1}
 &\displaystyle\max_{\xi,\;\mathbf{S}} \xi\\
    \text{  s.t.  }\nonumber\\
    &tr(\mathbf{X_uS})+|\rho_{u0}[l+n]\sqrt{\beta_{u0}}^{-1}\hat{h}_{u0}|^2\geq\xi\;\forall u\in\mathcal{U}\\
    & \mathbf{S}_{m,m}=1\;1\leq m\leq M+1\\
    &\mathbf{S}\succeq 0\\
    &\xi\geq 0
\end{align}


\noindent Here $tr(.)$ denotes the trace of a matrix.  The SDP in \eqref{SDR1} can easily be solved using CVX \citep{CVX} followed by Gaussian randomization step to obtain the rank $1$ optimal solution  \citep{OFDMA_embb_urllc}, \citep{D2D_ris}.

\subsection{Power allocation}

After determining the phase shifts of the RIS, we consider an optimal power allocation problem to maximize the aggregate  throughput achieved by all the UEs. For a given time slot $l+n$, we need to find the power allocation to all UEs such that both $R(l+n)$ and $\bar R(l+n)$ are maximized. Let $\mathbf{P}_n=[P_{1n},\;P_{2n},\;\cdots,\;P_{Un}]$ denote the transmit power vector for $U$ UEs at time slot $l+n$ to compute $R(l+n)$.
Similarly, let $\mathbf{\bar P}_n=[\bar P_{1n},\;\bar P_{2n},\;\cdots,\;\bar P_{Un}]$ denote the transmit power vector for $U$ UEs at time slot $l+n$ to compute $\bar R(l+n)$. By $P_{un}$ and $\bar{P_{un}}$, we mean the power allocated to UE $u$ at time slot $l+n$ for actual and expected throughput respectively. In order to compute the aggregate throughput gain,  we need to optimize both $\mathbf{P}_n$ and  $\mathbf{\bar P}_n$. This is because, the gain is measured as the difference between the maximum achievable throughput without CE and the maximum expected throughput had CE been performed at time slot $l+n$. Now, to formulate the optimization problems, we define two auxiliary variable vectors $\boldsymbol\eta_n=[\eta_{1n},\;\eta_{2n}\;\cdots,\;\eta_{Un}]$ 
and $\boldsymbol{\bar{\eta}}_n=
[\bar{\eta}_{1n},\;\bar{\eta}_{2n},\;\cdots,\;\bar{\eta}_{Un}]$. Furthermore, denoting $R^{(1)}=R,\;\eta^{(1)}=\eta,\;R^{(2)}=\bar R$, $\mathbf{P}^{(1)}=\mathbf{P}$, $\mathbf{P}^{(2)}=\mathbf{\bar P}$ and $\eta^{(2)}=\bar\eta$ the two throughput optimization problems can be formulated as follows:
\begin{align}\label{opt1}
    &\displaystyle\max_{\boldsymbol{\eta_n}}\sum_{u=1}^U\eta_{un}^{(.)}\\
    \text{ s.t. }\\
    &\eta_{un}^{(.)}-bR_u^{(.)}(l+n)\leq0,\; u\in\mathcal{U}\label{c2}\\
    &\sum_{u=1}^U bP_{un}^{(.)}\leq P_{tot} \label{c3}\\
    &P_{un}^{(.)}\geq0,\;u\in\mathcal{U}\label{c4} \\ 
    &\eta_{un}^{(.)}\geq0,\;u\in\mathcal{U}
    \end{align}





It is to be noted that $R_u(l+n) $ and $\bar  R_u(l+n)$ are concave function in $\mathbf{P}_{n}$ and $\mathbf{\bar P}_n$ respectively. Therefore the optimization problem \eqref{opt1} is separable convex optimization problems. Moreover, it may be easily observed that the constraint \eqref{c2} and \eqref{c3} satisfy the Slater's condition. Therefore the convex optimization problems follows strong duality. We solve the dual of the above convex optimization to obtain a closed form expression for $P_{un}$ and $\bar P_{un}$. For ease of representation, we use the following notations: 
\begin{align*}
&A^{(1)}_u=|\rho_u[l+n]\hat{\mathbf{g}}_u\mathbf{L}_u\mathbf{\Phi}\mathbf{h}+\rho_{u0}[l+n]\sqrt{\beta_{u0}^{-1}}\hat{h}_{u0}|^2\\
 &A^{(2)}_u=\frac{M\rho_u^2}{\beta_u} + \frac{M(M-1)(\rho_u[l+n]\Omega_1\Omega_2)^2}{\beta_u}+\frac{2M\rho_u[l+n]\rho_{u0}[l+n]\Omega_0\Omega_1\Omega_2}{\sqrt{\beta_{u0}\beta_u}}\\
&+\rho_{u0}[l+n]^2\beta_{u0}^{-1}\\
&B_u=\frac{ M \bar{\rho_u[l+n]}^2\left(1-\Omega_2^2\right)}{\beta_u}+\frac{ \bar{\rho_{u0}[l+n]}^2\left(1-\Omega_0^2\right)}{\beta_{u0}}\\ &C_u=\sigma_u^2+\exp(2\mu^u+2(\Sigma^u)^2). 
\end{align*}
Therefore, $bR(l+n)$ and $b\bar R(l+n)$ can be computed as:

\begin{eqnarray}
& bR^{(1)}(l+n)=bR(l+n)=   b\left(1-\frac{\alpha_{l+n}T_p}{N_c}\right)W\log_2\left(1+\frac{A^{(1)}_uP_{un}}{B_uP_{un}+C_u}\right)= \nonumber\\
 &W^{(1)} .\log_2\left(1+\frac{A^{(1)}_uP_{un}}{B_uP_{un}+C_u}\right),\\
 &bR^{(2)}(l+n)=b\bar R(l+n)=    b\left(1-\frac{\bar\alpha_{l+n}T_p}{N_c}\right)W\log_2\left(1+\frac{ A^{(2)}_uP_{un}}{B_uP_{un}+C_u}\right)=\nonumber\\
 &W^{(2)} \log_2\left(1+\frac{ A^{(2)}_uP_{un}}{B_uP_{un}+C_u}\right).
\end{eqnarray}
\noindent Here, $W^{(1)}=b\left(1-\frac{\alpha_{l+n}T_p}{N_c}\right)W$ and $W^{(2)}=b\left(1-\frac{\bar\alpha_{l+n}T_p}{N_c}\right)W$. The Lagrangian for the objective function \eqref{opt1} and the constraints \eqref{c2} and \eqref{c3} is as follows:

\begin{align}\label{lag}
    \mathcal{L}^{(.)}(\boldsymbol{\eta}^{(.)},\mathbf{P}_n^{(.)},\boldsymbol{\zeta}^{(.)},\nu^{(.)})=\sum_{u=1}^U\eta^{(.)}_{un}+\sum_{u=1}^U \zeta_u^{(.)}\left(bR_u^{(.)}(l+n)-\eta_{un}^{(.)}\right)\nonumber\\
    +\nu^{(.)}\left(P_{tot}-\sum_{u=1}^U P^{(.)}_{un}\right).
\end{align}
\noindent Here, $\boldsymbol{\zeta}^{(.)}=[\zeta_{1},\;\zeta_2,\;\cdots,\;\zeta_U]$ and $\nu^{(.)}$ are the Lagrangian dual variables. Therefore, the Lagrangian dual of \eqref{opt1} is formulated as follows:

\begin{align}
    \displaystyle\min_{\nu>0,\;\boldsymbol{\zeta}}\mathcal{G}(\nu,\boldsymbol{\zeta})=\sup_{\boldsymbol{\eta}^{(.)} ,\mathbf{P}^{(.)}}\mathcal{L}^{(.)}(\boldsymbol{\eta}^{(.)},\mathbf{P}_n^{(.)},\boldsymbol{\zeta}^{(.)},\nu^{(.)}),
\end{align}
\noindent where $\mathcal{G}(\nu,\boldsymbol{\zeta})=\displaystyle\inf_{\boldsymbol{\eta}^{(.)} ,\mathbf{P}^{(.)}}\mathcal{L}^{(.)}(\boldsymbol{\eta}^{(.)},\mathbf{P}_n^{(.)},\boldsymbol{\zeta}^{(.)},\nu^{(.)})$ is the Lagrangian dual objective. Now, \eqref{lag} is convex, continuous and differentiable in $\mathbf{P}^{(.)}_n$ and $\boldsymbol{\eta}^{(.)}$. The critical points of the Lagrangian can be computed from the following equations:

\begin{eqnarray}
     &\frac{\partial \mathcal{L}^{(.)}}{\partial \eta^{(.)}_{un}}=0,\;\forall u\in\mathcal{U}\label{cr1}\\
      &\frac{\partial \mathcal{L}^{(.)}}{\partial P_{un}^{(.)}}=0,\;\forall u\in\mathcal{U}\label{cr2}
\end{eqnarray}
From \eqref{cr1} we compute the values of $\zeta_u$ $\forall u\in \mathcal{U}$ as follows:
\begin{align*}
    &\frac{\partial \mathcal{L}^{(.)}}{\partial \eta^{(.)}_{un}}=0,\;\forall u\in\mathcal{U}\\
    &\implies 1-\zeta_u=0,\;\forall u\in\mathcal{U}\\
    &\implies \zeta_u=1,\;\forall u\in\mathcal{U}
\end{align*}
\noindent Substituting $\zeta_{un}$ in \eqref{lag} and for a given value of $\nu$, the closed form expression of $P^{(.)}_{un}\;\forall u\in\mathcal{U}$ is obtained from \eqref{cr2} as follows: 

\begin{align}
   &\frac{\partial\mathcal{L}^{(.)}()}{\partial P_{un}}=0,\nonumber\\
   &\implies (A^{(.)}_u+B_u)B_uP_{un}^2 + P_{un}C_u(A^{(.)}_u+2B_u)+\nonumber\\
   &C_u^2-\frac{W^{(.)}A^{(.)}_uC_u}{\nu^{(.)}}=0.\nonumber\\
\end{align}
\noindent Considering $a_u=(A^{(.)}_u+B_u)B_u$, $b_u=C_u(A^{(.)}_u+2B_u)$ and discarding the negative root, we find the value of $P^{(.)}_{un}$ as follows:

\begin{equation}\label{p_value}
    P^{(.)\nu}_{un}=\frac{-b_u+\sqrt{b^2-4a(C_u^2-\frac{W^{(.)}A^{(.)}_uC_u}{\nu^{(.)}})}}{2a}.
\end{equation}
\noindent Therefore, for $\nu^{(.)*}=\text{\rm argmin }\mathcal{G}(\nu^{(.)})$
we obtain the the optimal power allocated for UE $u$ at time slot $l+n$ denoted by $P^{(.)*}_{un}$ is obtained as follows:

\begin{equation}
    P^{(.)*}_{un}=\frac{-b+\sqrt{b^2-4a(C_u^2-\frac{W^{(.)}A^{(.)}_uC_u}{\nu^{(.)*}})}}{2a}.
\end{equation}
\noindent Furthermore, $\mathcal{G}(\nu^{(.)})$ is convex in $\nu^{(.)}$ since it is the point-wise supremum of an affine function. Hence, we can obtain $\nu^{(.)*}$ iteratively by simple gradient descent after substituting $P^{(.)}_{un}$ from \eqref{p_value} in \eqref{lag}. The entire optimization process has been formally described in Algorithm \ref{optalgo}.

\begin{algorithm}
    \caption{Power allocation at slot ($l+n$)} \label{optalgo}
    \KwData{Gradient descent step $\chi$, Acceptable error margin $\epsilon$}
    \KwResult{Optimal power allocation $\mathbf{P}^{(.)*}_n$}
    Initialize $i=0$, $\nu^{(.)i}=100$, $\mathbf{P}_n=\mathbf{0}$\;
    Compute $\mathbf{P}_n^{(i)}$ according to \eqref{p_value} for $\nu^{i}$\;\label{l2}
    \For{$P^{(.)i}_{un}\in\mathbf{P}^{(.)i}_n$}{ \If {$P^{(.)i}_{un}\leq 0$}{Set $P^{(.)i}_{un}=0$\;}}
    Substitute $\mathbf{P}^{(.)i}$ in \eqref{lag}\;
    $\nu^{(.)i+1}=\left(\nu^{(.)i}-\chi\left(\frac{\partial\mathcal{L}(\mathbf{P}^{(.)i}_n,\nu)}{\partial \nu}|_{\nu=\nu^{i}}\right)^+\right)^+$\;
    Compute $\mathbf{P}_n^{(.)i+1}$ according to \eqref{p_value} for $\nu^{i+1}$\;
    \For{$P^{(.)i+1}_{un}\in\mathbf{P}^{(.)i+1}_n$}{ \If {$P^{(.)i+1}_{un}\leq 0$}{Set $P^{(.)i+1}_{un}=0$\;}}
    \If{$|\mathbf{P}_n^{(.)i+1}-\mathbf{P}_n^{(.)i}|\leq\epsilon$}{$\mathbf{P}^{(.)*}_n=\mathbf{P}_n^{(.)i+1}$\;}
    \Else{$i=i+1$ and go to line \ref{l2}\;}
    \textbf{Return} $\mathbf{P}^{(.)*}_n$
\end{algorithm}

\subsection{Algorithm for skipping CE phase}
 
Considering that the last CE was done at time slot $l$ we need to find the number of consecutive time slots where CE can be skipped such that the aggregate throughput gain is maximised. For each $n\in\mathcal{T}$ the first step is to compute the RIS phase shifts by solving the optimization problem \eqref{SDR1}. The next step is to compute the throughput gain achieved by each UE if it decides to skip CE in the time slot $l+n$. For $1\leq n\leq \lfloor{\frac{\Psi}{T_c}}\rfloor$, it is to be decided weather the UEs should communicate without any CSI acquisition for the entire duration of $\Psi$ or CE is necessary before $\Psi$ ends. Before computing the aggregate throughput gain, we first obtain the expected throughput if the CE was performed at the beginning of time slot $l+n$ by computing the optimal power allocation that maximises $\displaystyle\sum_{u\in\mathcal{U}}\bar R_u (l)$ following Algorithm \ref{optalgo}. Then we compute the optimal power allocation to the UEs such that $\displaystyle\sum_{u\in\mathcal{U}}R_u(l+n)$ is maximised following Algorithm \ref{optalgo}. Finally the aggregate throughput gain is computed using \eqref{gain} and \eqref{sumgain}. Next, for $\lfloor\frac{\Psi}{T_c}\rfloor+1\leq n\leq N_{max}-1$,  it is to be decided weather to continue communication without CSI acquisition in time slot $l+n$ or perform CE. Following Algorithm \ref{optalgo} we compute optimal power allocation to maximise $\displaystyle\sum_{u\in\mathcal{U}}\bar R_u (l+ n \mod \lfloor\frac{\Psi}{T_c}\rfloor)$ and $\displaystyle\sum_{u\in\mathcal{U}} R_u (l+n)$. Finally we calculate the aggregate throughput gain using \eqref{gain} and \eqref{sumgain}. 
After computing the aggregate throughput gain, we check the following two conditions:
\begin{enumerate}
    \item Weather $G_n>0$.
    \item Weather $R_u(l+n)>R^{th}$ for all $u\in\mathcal{U}$.
\end{enumerate}
\noindent The process terminates upon either encountering the first $n\in\mathcal{T}$ where the above cited condition is violated or when $n=N_{max}-1$. The algorithm is formally described as follows in Algorithm \ref{algo1}.   

\begin{algorithm}
    \caption{CE skipping prediction algorithm}\label{algo1}
    \KwData{Time slot $l$ when CE was last performed}
    \KwResult{S, the number of consecutive time slots where CE can be skipped}
    Set $S=0$\;
    \For{$n\in\mathcal{T}$}{
    \If{$1\leq n\leq \lfloor{\frac{\Psi}{T_c}}\rfloor$}{
    Compute $\mathbf{P}^{(1)*}_n$ for slot $l+n$ and $\mathbf{P}^{(2)*}_n$ for slot $l$ using Algorithm \ref{optalgo}\;}
    \Else{Compute $\mathbf{P}^{(1)*}_n$ and $\mathbf{P}^{(2)*}_n$ for slot $l+n$ using Algorithm \ref{optalgo}\;}
    Compute gain $G_n$ from \eqref{sumgain}\;
    \If{$G_{n}>0$ and $R_u(l+n)>R^{th} \; \forall u\in\mathcal{U}$ }{
    $S=S+1$\;
    Continue \;
    }
    \Else{
    Break\;
    }
    }
    \textbf{Return} $S$
\end{algorithm}

\section{Numerical results}\label{sim}

In this section, we evaluate the performance of our proposed algorithm through extensive system-level simulations. We compare the scenario where CSI estimation is initiated based on our proposed algorithm with the following scenarios used by recent studies such as \citep{RIS_UAV_ergodic_throughput_max}, \citep{RIS_outdate_snr} and \citep{UAV_RIS_selection} :

\begin{itemize}
    \item \textbf{Comparison basis 1:} $\Psi=\frac{c\cos^{-1}(0.9)}{2\pi Vf_{c}}$. Here $V$ is the maximum velocity achieved by any UE  \citep{lower_bound_coherence_time}.
    \item \textbf{Comparison basis 2:} $\Psi=\frac{c\cos^{-1}(0.7)}{2\pi Vf_{c}}$. Here $V$ is the maximum velocity achieved by any UE  \citep{lower_bound_coherence_time}.
\end{itemize}
The CE process itself is performed using algorithm in \citep{CE_Multi_UE_RIS} and \citep{CE_Multi_UE_RIS_framework} with pilot overhead $T_p= M+1 $.
 Through simulations, we show the impact of UE velocity, number of UEs and number of RIS elements on the sum throughput of the network.  
\subsection{Simulation setup}

\begin{table}[htbp]
\footnotesize
    \centering
    \caption{Simulation parameters}
    \begin{tabular}{|c|c|}
    \hline
        \textbf{Parameters} & \textbf{values} \\\hline
         Radius of gNB coverage & $100$ meters\\\hline
         Velocity range & $30$ m/s to $70$ m/s\\\hline
         gNB coordinate & $(50,0,10)$\\\hline
         RIS coordinate & $(25,30,10)$ \\\hline
         No. of RIS elements & $200$ to $500$ \\\hline
         $P_{tot}$ & $20$ dBm\\\hline
         Carrier frequency & $5$ Ghz\\\hline
         $T_c$ & 0.06 seconds\\\hline
         Noise power & $-120$ dBm\\\hline
         \begin{tabular}{@{}c@{}}Pathloss exponent\\ gNB-RIS-UE channel\end{tabular}&$2$\\\hline
        \begin{tabular}{@{}c@{}} Pathloss exponent\\direct channel\end{tabular} & $3.5$\\\hline
         Rician factors & $K_1=K_2=1$ dB, $K_0=0$\\\hline
         Bandwidth per RB & $180$ KHz\\\hline
         No. of RBs & $96$ \\\hline
         Background interference & $-95$ dBm \\\hline
         $\tau_{max}$ & $250$ ns \\\hline
         $R^{th}$ & $256$ Mbps\\\hline
         Total communication time & $50$ seconds\\\hline

    \end{tabular}
    
    \label{sim_parm}
\end{table}

The parameters of the simulation environment have been adopted from \citep{tong2022two_stage}, \citep{OFDMA_embb_urllc}, \citep{RIS_outdate_snr}, \citep{lower_bound_coherence_time} and \citep{RIS-performance}. The UEs are considered to move in a circular region of radius $100$ meters. The centre of the circular region is considered to be the origin. The initial positions of the UEs are chosen uniformly randomly inside the circular region. The UEs move linearly in arbitrary directions with a velocity ranging from $30$ m/s to $70$ m/s throughout the simulation time. The gNB is placed at $(50,0,10)$ and the coordinate of the RIS centre is $(25,30,10)$. All distances are in meters. The total allowable transmission power of the gNB is set at $P_{tot}=20$ dBm \citep{RIS_outdate_snr}. The carrier frequency is set at $5$ GHz \citep{RIS_outdate_snr}. We compute $T_c$ to be $0.06$ seconds by considering $\rho^{th}=0.9$. The noise power is $-120$ dBm. Moreover, the gNB and UE antenna gains are $20$ dB and $0$ dB respectively \citep{RIS_outdate_snr}. The pathloss exponent of the cascaded gNB-RIS-UE channel is set at $2$, and the direct gNB-UE channel is set at $3.5$ \citep{RIS-performance}. We set the Rician factors as $K_1=K_2=1$ dB and $K_0=0$ \citep{RIS-performance}. The  bandwidth per RB is set as $180$ KHz \citep{OFDMA_embb_urllc}. Number of resource blocks is $96$ \citep{OFDMA_embb_urllc}. The background interference from neighbouring cells is set at a value of $-95$ dBm \citep{tong2022two_stage}. The default number of RIS elements is $M=300$. The maximum delay spread is set at $\tau_{max}=250$ ns \citep{lower_bound_coherence_time}. The minimum required data rate is set at $R^{th}=256$ Mbps. Finally, the total time of communication is $50$ seconds.  

\subsection{Evaluating aggregate throughput}

\begin{figure}[h!]

    \centering
    \includegraphics[scale=.31]{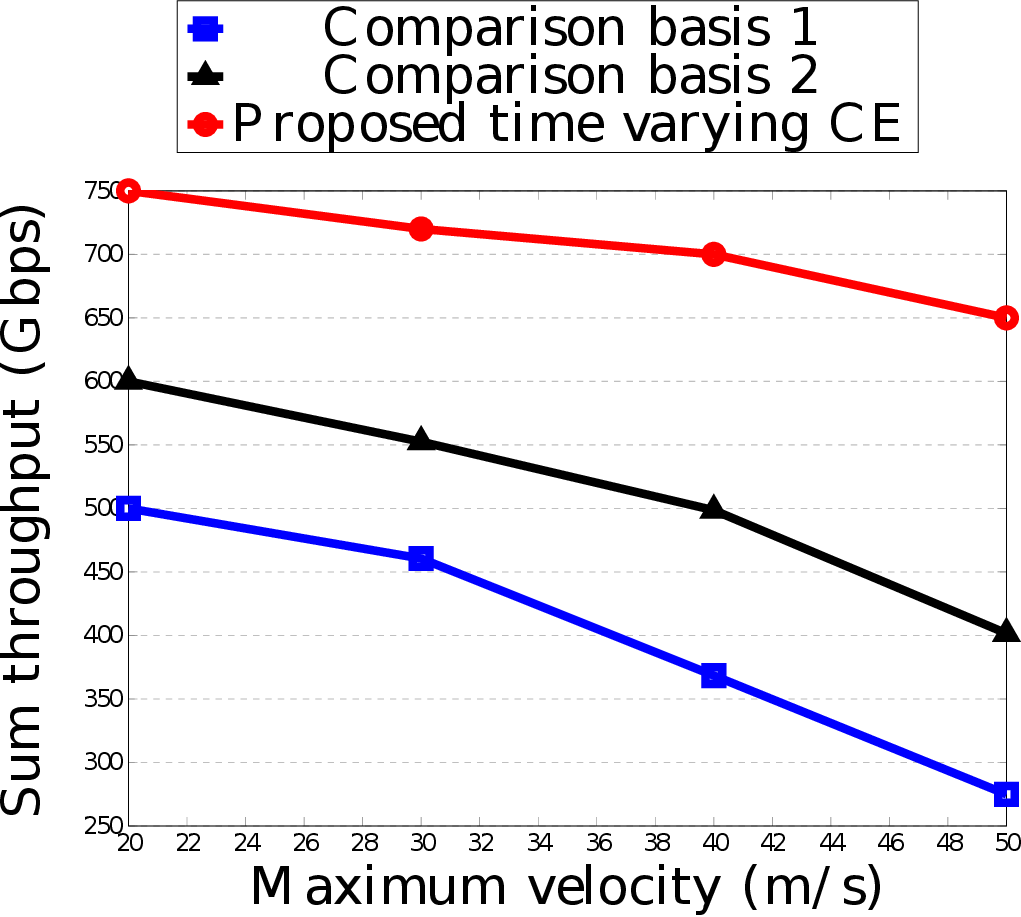}
    \caption{Throughput vs. velocity (Number of elements=300, Number of devices=20)}
    \label{rate_v}
  
\end{figure}

\begin{figure}[h!]
   \centering
        \includegraphics[scale=.31]{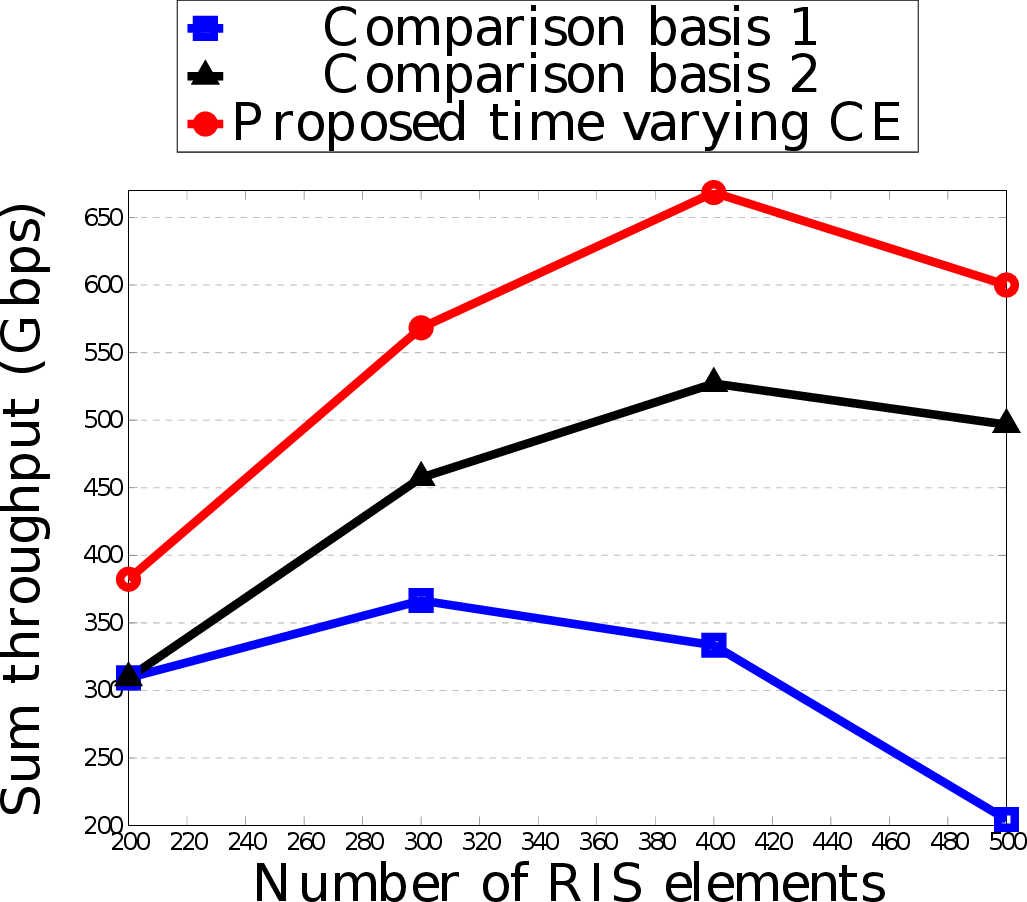}
    \caption{Throughput vs. Number of elements (Number of devices=20, velocity=40 m/s)}
    \label{rate_ele}
  
\end{figure}

\begin{figure}[h!]

\centering
   \includegraphics[scale=.31]{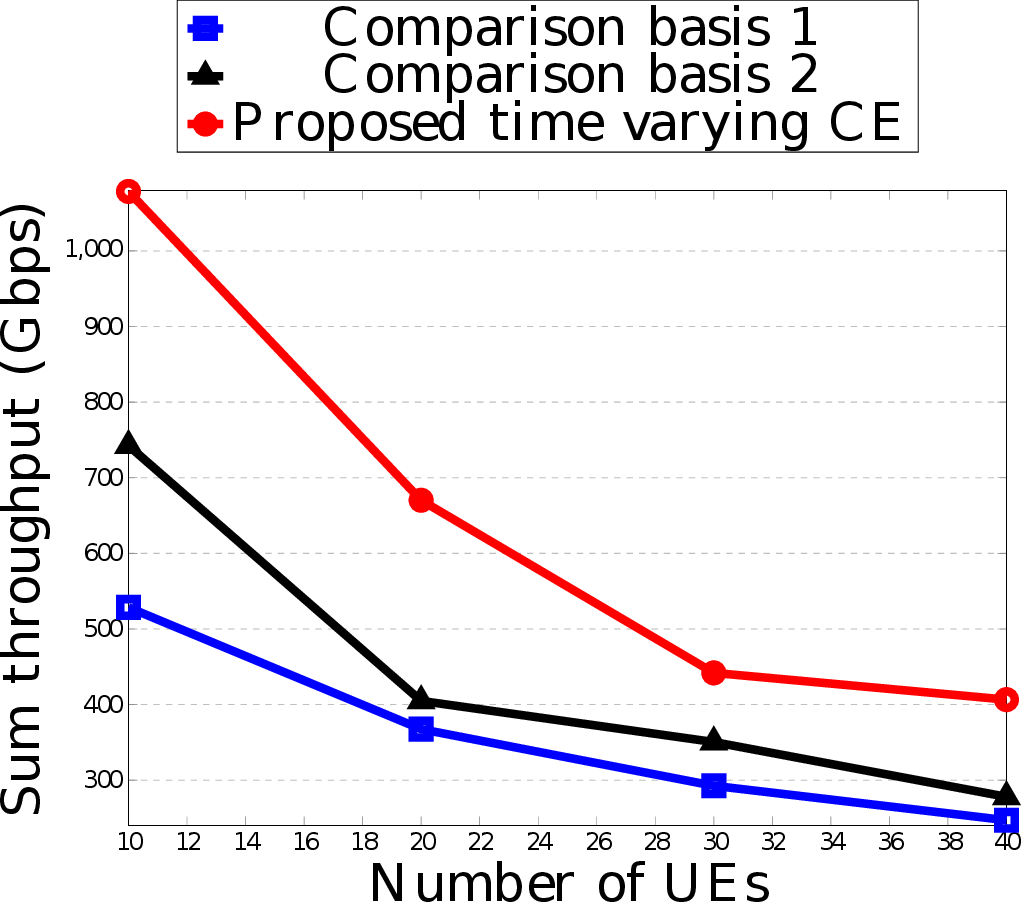}
    \caption{Throughput vs. Number of devices (Number of elements=300, velocity=40 m/s)}
    \label{rate_tl}

\end{figure}

Fig. \ref{rate_v} shows the effect of the velocity of UE on the sum throughput of the network for the scenario of CSI estimation initiation based on our proposed algorithm and that adopted in comparison basis 1 and 2. The number of UEs is $20$ and the number of RIS elements is $300$. The velocity ranges from $20$ m/s to $50$ m/s with a step of $10$ m/s. The throughput monotonically decreases for all three scenarios which can be explained as follows. From \eqref{timesloteq}, it can be observed that the channel coherence block length and the duration of a time slot decreases with increasing velocity. Furthermore, the rate at which the channel estimates become outdated is higher for UEs with higher velocity. As a consequence, the frequency of CSI acquisition increases with increasing maximum allowable velocity thereby increasing pilot overhead and reducing aggregate throughput. It can also be observed that the throughput for our proposed strategy is higher as compared to both comparison basis 1 and 2. Depending on the channel conditions and velocities of each UE, the duration between consecutive CSI acquisition is either shorter or longer than that in the case of the comparison bases. A longer duration between CE results in reduced pilot overhead as CSI acquisition is skipped in many consecutive time slots. On the other hand when the inter CE duration of the comparison bases 1 and 2 become too long and the channel estimates becomes degradingly outdated, our proposed algorithm performs CE more frequently. It is also observed that the gain in throughput for the scenario employing our proposed algorithm increases with increasing maximum allowable velocity. This is because the coherence time decreases with increasing velocity hence, the time left for actual data communication before the next CE in the case of comparison bases decreases with increasing velocity. On the other hand, our proposed time varying CE estimation algorithm dynamically skips the required number of time slots such that the aggregate throughput gain is maximized.  

Fig. \ref{rate_ele} shows the relationship between the number of RIS elements and the sum throughput of the network for the proposed algorithm and the comparison bases 1 and 2. The velocity of UEs is considered to be  $40$ m/s and the number UEs is $30$. The number of RIS elements ranges from $200$ to $500$ with a step of $100$ RIS elements. It is observed that the sum throughput initially increases with the number of RIS elements upto a threshold value $400$ for proposed algorithm and comparison basis 2 and $300$ for the case of comparison basis 1. This is because the throughput achieved by each UE increases with increasing number of RIS elements. However, the pilot overhead also increases linearly with the number of elements such that after a certain point ($400$ for proposed algorithm, $400$ for comparison basis 2 and $300$ for comparison basis 1) the pilot overhead consumes s significant portion of the time between consecutive CE thus reducing the throughput. It may be observed that for all values of the number of RIS elements shown in the result, the aggregate throughput for the proposed algorithm is higher than that for the comparison bases 1 and 2.  

Fig. \ref{rate_tl} shows the relationship between the number of UEs served by the gNB and the sum throughput of the network for the scenario of CSI estimation initiation based on our proposed algorithm and comparison bases 1 and 2. The velocity of UEs is considered to be  $40$ m/s and the number of RIS elements is $300$. The number of UEs ranges from $10$ to $40$ with a step of $10$ UEs. It is observed that the throughput for the scenario employing our proposed strategy is higher compared to both comparison bases. This is because our proposed algorithm suitably changes the interval between consecutive CE phases based on reasons already discussed for the results in Fig. \ref{rate_v}. Moreover, it is observed that throughput monotonically decreases beyond $20$ UEs. This is because with the increasing number of UEs the power share for each UE decreases.



\subsection{Evaluating variation in the inter-CE interval}

\begin{figure}
 \begin{minipage}[t]{0.5\textwidth}
    \centering
    \includegraphics[scale=0.5]{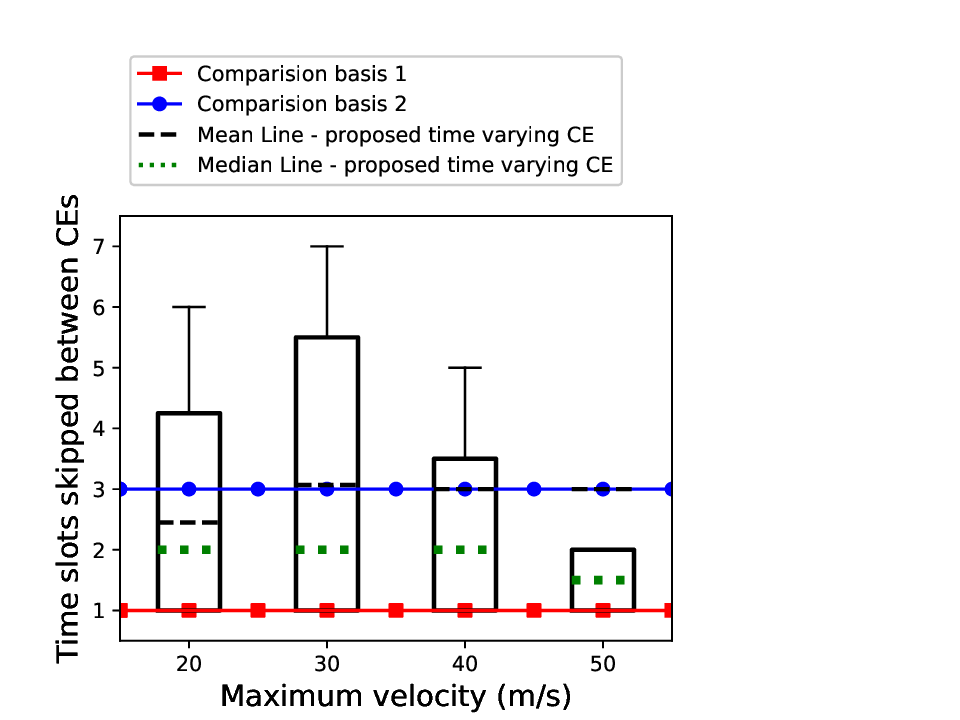}
    \caption{Time slots skipped between consecutive CE vs. velocity (Number of elements=300, Number of devices=20)}
    \label{vslot}
    \end{minipage}\hfill
    \begin{minipage}[t]{0.5\textwidth}
     \centering
    \includegraphics[scale=.31]{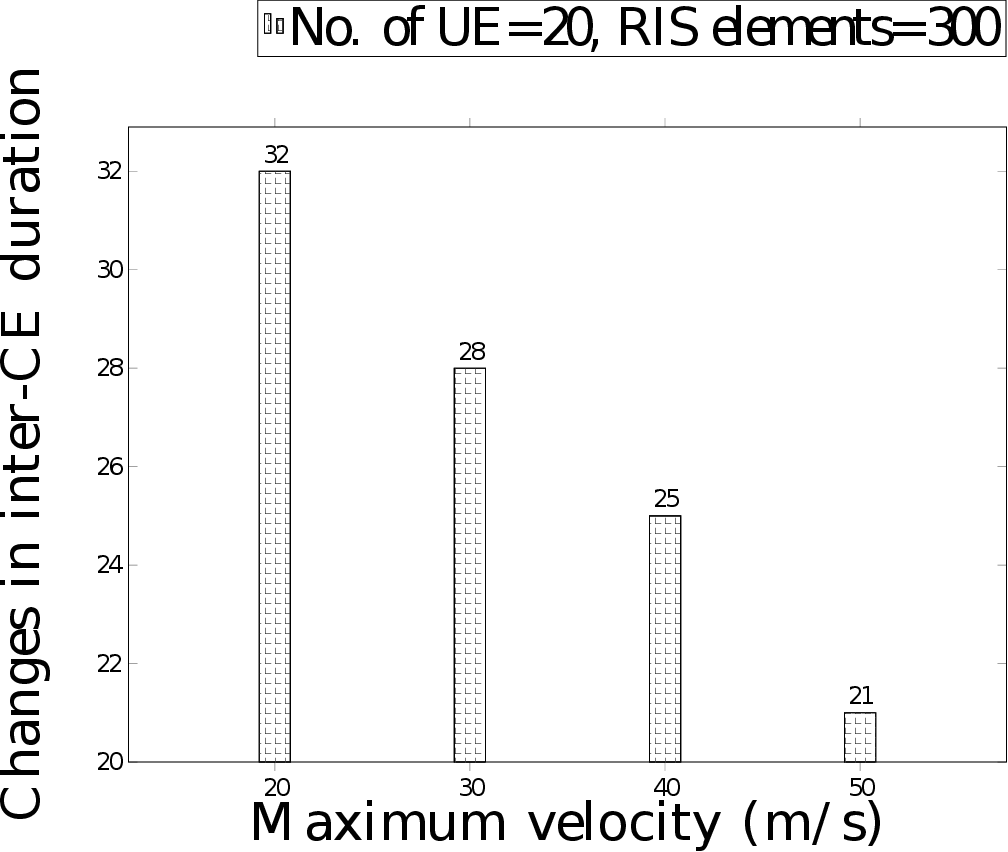}
    \caption{Frequency of change in inter-CE duration vs. velocity (Number of elements=300, Number of devices=20)}
    \label{f_v}    
    \end{minipage}\hfill
\end{figure}

In this subsection, we evaluate the dynamic nature of the time interval between two consecutive CE phases. Fig. \ref{vslot} is a box plot that shows the distribution of the number of time slots skipped between two consecutive CE phases during the time of communication for different values of maximum allowable velocities. The number of UEs is $20$ and the number of RIS elements is $300$. The velocity ranges from $20$ m/s to $50$ m/s with a step of $10$ m/s. It can be observed that the number of time slots that are skipped according to our proposed algorithm varies with time to maximize the aggregate throughput as opposed to the case for comparison bases 1 and 2. The range of this variation initially increases upto a maximum allowable velocity of $30$ m/s and then decreases with further increase in the value of velocity. Initially, the range between the lowest number of slots skipped and the highest number of slots skipped increases as the range of velocities of different UEs increases. This is because depending on the position and channel condition of different UEs sometimes a longer duration between two consecutive CEs becomes optimal. However, this range decreases with further rise in velocity beyond $30$ m/s. This is because as some UEs are in very high velocities their channel estimates become detrimentally outdated much faster compared to the rest of the UEs. This forces frequent CSI acquisition. Fig. \ref{f_v} shows the relationship between maximum allowable velocity and the number of times the duration between consecutive CE phases changes throughout the communication period. It can be observed that the frequency of change in the duration between consecutive CE phases decreases monotonically with velocity. This is because with higher velocity the time slots skipped between consecutive CE is reduced and the value of the number of slots skipped is concentrated near its median value as shown in Fig. \ref{vslot}.

\begin{figure}
 \begin{minipage}[t]{0.5\textwidth}
    \centering
    \includegraphics[scale=0.5]{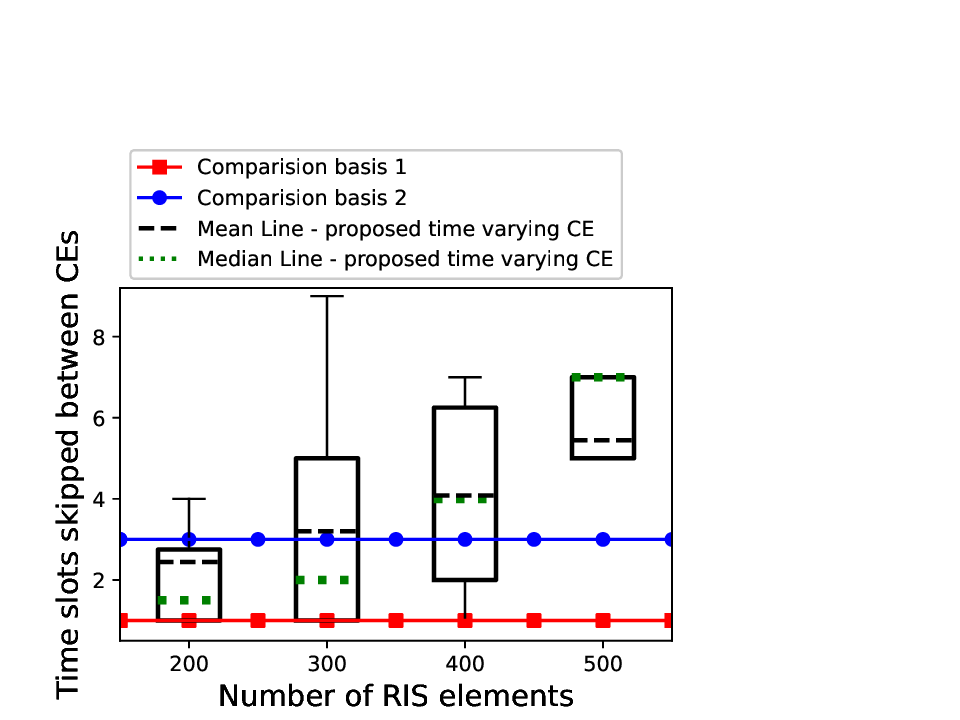}
    \caption{Time slots skipped between consecutive CE vs. number of RIS elements (Number of UEs=30, Maximum velocity=40 m/s)}
    \label{rslot}
    \end{minipage}\hfill
     \begin{minipage}[t]{0.5\textwidth}
        \includegraphics[scale=.31]{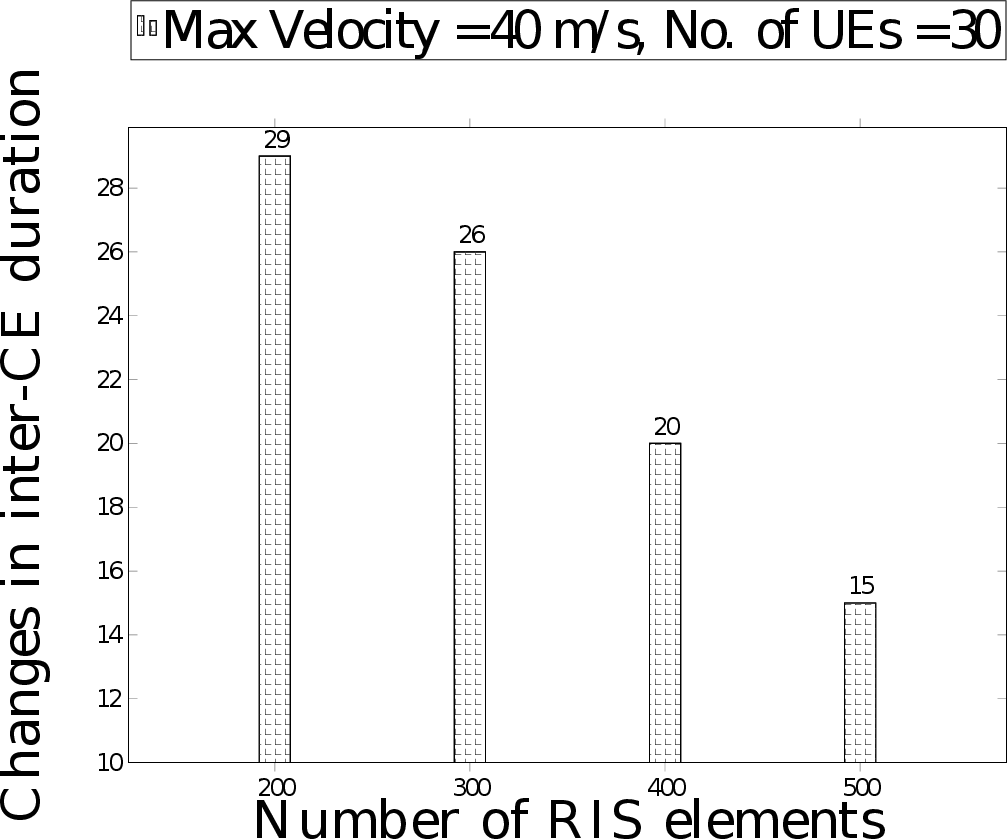}
    \caption{Frequency of change in inter-CE duration vs. Number of elements (Number of devices=30, velocity=40 m/s)}
    \label{f_ele}
    \end{minipage}
\end{figure}

Fig. \ref{rslot} is a box plot that shows the distribution of the number of time slots skipped between two consecutive CE phases during the time of communication for different number of RIS elements. The velocity of UEs is considered to be  $40$ m/s and the number UEs is $30$. The number of RIS elements ranges from $200$ to $500$ with a step of $100$ RIS elements. Initially the spread in the data representing the number of slots skipped between consecutive CEs increases upto a threshold number of RIS elements ($400$). Beyond $400$ elements, the spread decreases. Furthermore, the median of the data monotonically increases with the number of RIS elements. This can be explained as follows. Initially, low number of RIS elements result in poor channel gain at the UE thereby requiring frequent CE phases to prevent the channel from becoming degradingly outdated. With an increase in the number of RIS elements the channel gain at each UE increases which allows for greater delay tolerance before channel needs to be estimated again. However, the pilot overhead also rises with the increasing number of elements. Therefore, for high number of RIS elements (i.e beyond $400$ elements in this case), to overcome the loss in data communication time due to high pilot overhead, the distribution of the number of time slots skipped in concentrated around a high median value of $7$. Fig. \ref{f_ele} shows the relationship between number of RIS elements and the number of times the duration between consecutive CE phases changes throughout the communication period. It can be observed that the frequency of change in the duration between consecutive CE phases decreases monotonically with the increasing  number of RIS elements. This is because higher number of elements leads to higher pilot overhead. Therefore as shown in Fig. \ref{rslot} the time slots skipped between consecutive CE is increased and the value  of the number of slots skipped is concentrated near its median value.
\begin{figure}[h!]
 \begin{minipage}{0.5\textwidth}
    \centering
    \includegraphics[scale=0.5]{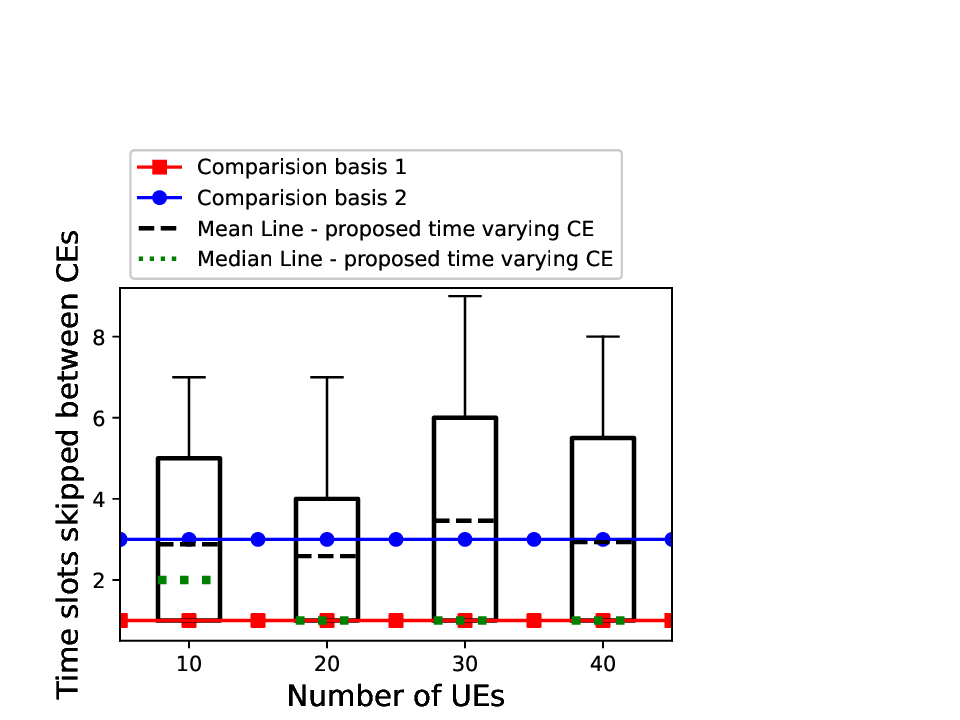}
    \caption{Time slots skipped between consecutive CE vs. number of UEs (Number of elements=300, Maximum velocity=40 m/s)}
    \label{uslot}
    \end{minipage}\hfill
    \begin{minipage}{0.5\textwidth}
    \includegraphics[scale=.31]{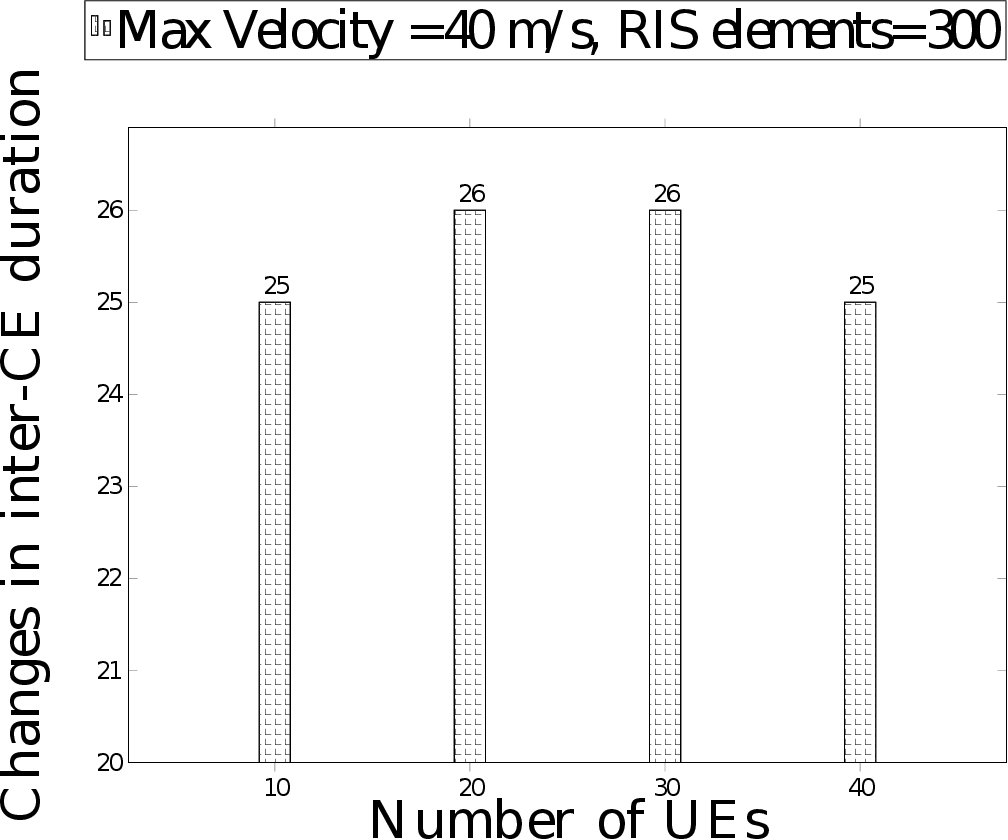}
    \caption{Frequency of change in inter-CE duration vs. Number of devices (Number of elements=300, velocity=40 m/s)}
    \label{f_tl}
    \end{minipage}\hfill
\end{figure}

Fig. \ref{uslot} is a box plot that shows the distribution of the number of time slots skipped between two consecutive CE phases during the time of communication for different number UEs present in the network. The velocity of UEs is considered to be  $40$ m/s and the number of RIS elements is $300$. The number of UEs ranges from $10$ to $40$ with a step of $10$ UEs. It can be observed that the number of time slots that are skipped according to our proposed algorithm varies with time to maximize the aggregate throughput as opposed to the case for comparison bases 1 and 2. The number of time slots needed to be skipped between consecutive CE primarily depends on the channel conditions and velocities of the UEs as reflected in Fig. \ref{vslot} and Fig. \ref{rslot}. As a result the number of UEs has no significant impact on the variation in the number of time slots skipped over time. Fig. \ref{f_tl} shows the relationship between number of UEs in the network and the number of times the duration between consecutive CE phases changes throughout the communication period. As verified by the results in Fig. \ref{uslot} the number of UEs in the network does not have a significant impact on the frequency of change in the duration between two consecutive CE.

\section{Conclusion}\label{con}
This manuscript tries to answer the question of when to conduct CE given outdated channel estimates in a RIS assisted wireless network. An algorithm is proposed in order to predict the number of time slots where CE need not be performed to utilize the time for sending actual data. The proposed algorithm takes into consideration the velocities and channel conditions of different UEs in the network and first computes optimal RIS phase shifts and power allocation for each UE to maximise the aggregate throughput. Finally the above computed phase shift optimization and power allocation is used to determine the aggregate throughput gain achieved at a given time slot where CE is skipped. Based on this gain value the number of time slots where CE need not be performed in order to maximise the sum throughput is predicted by the proposed algorithm. Extensive system level simulations show that our proposed time varying CE achieves higher throughput in comparison to coherence time based fixed CE interval in comparison bases 1 and 2.

\section{Author contributions}
\textbf{Souvik Deb:} Methodology, Software, Writing – original draft ; 
\textbf{Sasthi C. Ghosh:} Conceptualization, Writing – review \& editing;

\bibliographystyle{elsarticle-num}
\bibliography{ref}

\begin{thebibliography}{10}
\expandafter\ifx\csname url\endcsname\relax
  \def\url#1{\texttt{#1}}\fi
\expandafter\ifx\csname urlprefix\endcsname\relax\def\urlprefix{URL }\fi
\expandafter\ifx\csname href\endcsname\relax
  \def\href#1#2{#2} \def\path#1{#1}\fi

\bibitem{RIS_model}
M.~Jung, W.~Saad, M.~Debbah, C.~S. Hong, On the {O}ptimality of
  {R}econfigurable {I}ntelligent surfaces ({RIS}s): Passive {B}eamforming,
  {M}odulation, and {R}esource {A}llocation, \emph{IEEE Trans. Wireless
  Commun.} 20~(7) (2021) 4347--4363.
\newblock \href {https://doi.org/10.1109/TWC.2021.3058366}
  {\path{doi:10.1109/TWC.2021.3058366}}.

\bibitem{WPC_RIS_NOMA_select}
C.~Gong, X.~Dai, J.~Cui, K.~Long, Performance analysis of distributed
  reconfigurable intelligent surface aided noma systems, \emph{Wireless
  Personal Communications} (2023) 1--15\href
  {https://doi.org/https://doi.org/10.1007/s11277-023-10425-0}
  {\path{doi:https://doi.org/10.1007/s11277-023-10425-0}}.

\bibitem{modelling_RIS}
S.~Shen, B.~Clerckx, R.~Murch, Modeling and architecture design of
  reconfigurable intelligent surfaces using scattering parameter network
  analysis, IEEE Transactions on Wireless Communications 21~(2) (2022)
  1229--1243.
\newblock \href {https://doi.org/10.1109/TWC.2021.3103256}
  {\path{doi:10.1109/TWC.2021.3103256}}.

\bibitem{RIS_6G}
C.~Pan, H.~Ren, K.~Wang, J.~F. Kolb, M.~Elkashlan, M.~Chen, M.~Di~Renzo,
  Y.~Hao, J.~Wang, A.~L. Swindlehurst, X.~You, L.~Hanzo, Reconfigurable
  intelligent surfaces for 6g systems: Principles, applications, and research
  directions, IEEE Communications Magazine 59~(6) (2021) 14--20.
\newblock \href {https://doi.org/10.1109/MCOM.001.2001076}
  {\path{doi:10.1109/MCOM.001.2001076}}.

\bibitem{D2D_ris}
S.~Jia, X.~Yuan, Y.-C. Liang, Reconfigurable intelligent surfaces for energy
  efficiency in d2d communication network, IEEE Wireless Communications Letters
  10~(3) (2020) 683--687.

\bibitem{ris_mmwave}
H.~Du, J.~Zhang, J.~Cheng, B.~Ai, Millimeter wave communications with
  reconfigurable intelligent surfaces: Performance analysis and optimization,
  IEEE Transactions on Communications 69~(4) (2021) 2752--2768.

\bibitem{ris_vehicular}
M.~Alsenwi, M.~Abolhasan, J.~Lipman, Intelligent and reliable millimeter wave
  communications for ris-aided vehicular networks, IEEE Transactions on
  Intelligent Transportation Systems 23~(11) (2022) 21582--21592.
\newblock \href {https://doi.org/10.1109/TITS.2022.3190101}
  {\path{doi:10.1109/TITS.2022.3190101}}.

\bibitem{irs}
Q.~Wu, R.~Zhang, Towards {S}mart and {R}econfigurable {E}nvironment:
  {I}ntelligent {R}eflecting {S}urface {A}ided {W}ireless {N}etwork, \emph{IEEE
  Commun. Mag.} 58~(1) (2020) 106--112.
\newblock \href {https://doi.org/10.1109/MCOM.001.1900107}
  {\path{doi:10.1109/MCOM.001.1900107}}.

\bibitem{CE_framework}
A.~L. Swindlehurst, G.~Zhou, R.~Liu, C.~Pan, M.~Li, Channel estimation with
  reconfigurable intelligent surfaces—a general framework, Proceedings of the
  IEEE 110~(9) (2022) 1312--1338.

\bibitem{outdated_infocom}
S.~W. Haider~Shah, S.~Pavan~Deram, J.~Widmer, On the effective capacity of
  ris-enabled mmwave networks with outdated csi, in: IEEE INFOCOM 2023 - IEEE
  Conference on Computer Communications, 2023, pp. 1--10.
\newblock \href {https://doi.org/10.1109/INFOCOM53939.2023.10229028}
  {\path{doi:10.1109/INFOCOM53939.2023.10229028}}.

\bibitem{fast_ce_ris}
B.~Zheng, C.~You, R.~Zhang, Fast channel estimation for irs-assisted ofdm, IEEE
  Wireless Communications Letters 10~(3) (2021) 580--584.
\newblock \href {https://doi.org/10.1109/LWC.2020.3038434}
  {\path{doi:10.1109/LWC.2020.3038434}}.

\bibitem{ce_practicle_ris}
W.~Yang, H.~Li, M.~Li, Y.~Liu, Q.~Liu, Channel estimation for practical
  irs-assisted ofdm systems, in: 2021 IEEE Wireless Communications and
  Networking Conference Workshops (WCNCW), 2021, pp. 1--6.
\newblock \href {https://doi.org/10.1109/WCNCW49093.2021.9419982}
  {\path{doi:10.1109/WCNCW49093.2021.9419982}}.

\bibitem{ce_ris_mimo}
C.~Feng, W.~Shen, J.~An, L.~Hanzo, Joint hybrid and passive ris-assisted
  beamforming for mmwave mimo systems relying on dynamically configured
  subarrays, IEEE Internet of Things Journal 9~(15) (2022) 13913--13926.
\newblock \href {https://doi.org/10.1109/JIOT.2022.3142932}
  {\path{doi:10.1109/JIOT.2022.3142932}}.

\bibitem{RIS_outdate_snr}
Y.~Zhang, J.~Zhang, M.~Di~Renzo, H.~Xiao, B.~Ai, Reconfigurable intelligent
  surfaces with outdated channel state information: Centralized vs. distributed
  deployments, IEEE Transactions on Communications 70~(4) (2022) 2742--2756.
\newblock \href {https://doi.org/10.1109/TCOMM.2022.3146344}
  {\path{doi:10.1109/TCOMM.2022.3146344}}.

\bibitem{OFDMA_embb_urllc}
M.~Almekhlafi, M.~A. Arfaoui, M.~Elhattab, C.~Assi, A.~Ghrayeb, Joint
  scheduling of embb and urllc services in ris-aided downlink cellular
  networks, in: 2021 International Conference on Computer Communications and
  Networks (ICCCN), 2021, pp. 1--9.
\newblock \href {https://doi.org/10.1109/ICCCN52240.2021.9522196}
  {\path{doi:10.1109/ICCCN52240.2021.9522196}}.

\bibitem{optimal_group_strategy}
N.~K. Kundu, Z.~Li, J.~Rao, S.~Shen, M.~R. McKay, R.~Murch, Optimal grouping
  strategy for reconfigurable intelligent surface assisted wireless
  communications, IEEE Wireless Communications Letters 11~(5) (2022)
  1082--1086.

\bibitem{ris_grouping_q_learning}
S.~Tripathi, O.~J. Pandey, R.~M. Hegde, An optimal reflective elements grouping
  model for ris-assisted iot networks using q-learning, IEEE Transactions on
  Circuits and Systems II: Express Briefs 70~(8) (2023) 3214--3218.
\newblock \href {https://doi.org/10.1109/TCSII.2023.3251373}
  {\path{doi:10.1109/TCSII.2023.3251373}}.

\bibitem{performance_grouping}
Z.~Li, N.~K. Kundu, J.~Rao, S.~Shen, M.~R. McKay, R.~Murch, Performance
  analysis of ris-assisted communications with element grouping and spatial
  correlation, IEEE Wireless Communications Letters 12~(4) (2023) 630--634.

\bibitem{LIS_optimal_on/off}
N.~K. Kundu, M.~R. Mckay, Large intelligent surfaces with channel estimation
  overhead: Achievable rate and optimal configuration, IEEE Wireless
  Communications Letters 10~(5) (2021) 986--990.

\bibitem{csi_discrete_phase_shift}
C.~You, B.~Zheng, R.~Zhang, Channel estimation and passive beamforming for
  intelligent reflecting surface: Discrete phase shift and progressive
  refinement, IEEE Journal on Selected Areas in Communications 38~(11) (2020)
  2604--2620.

\bibitem{UAV_RIS_selection}
A.~Bansal, N.~Agrawal, K.~Singh, C.-P. Li, S.~Mumtaz, Ris selection scheme for
  uav-based multi-ris-aided multiuser downlink network with imperfect and
  outdated csi, IEEE Transactions on Communications (2023).

\bibitem{RIS_outdate_secrecyrate}
H.~Yang, Z.~Xiong, J.~Zhao, D.~Niyato, L.~Xiao, Q.~Wu, Deep reinforcement
  learning-based intelligent reflecting surface for secure wireless
  communications, IEEE Transactions on Wireless Communications 20~(1) (2021)
  375--388.
\newblock \href {https://doi.org/10.1109/TWC.2020.3024860}
  {\path{doi:10.1109/TWC.2020.3024860}}.

\bibitem{lower_bound_coherence_time}
R.~P. Torres, J.~R. P{\'e}rez, A lower bound for the coherence block length in
  mobile radio channels, Electronics 10~(4) (2021) 398.

\bibitem{time_varying_CSI}
M.~Xu, S.~Zhang, J.~Ma, O.~A. Dobre, Deep learning-based time-varying channel
  estimation for ris assisted communication, IEEE Communications Letters 26~(1)
  (2022) 94--98.
\newblock \href {https://doi.org/10.1109/LCOMM.2021.3127160}
  {\path{doi:10.1109/LCOMM.2021.3127160}}.

\bibitem{rappaport2024wireless}
T.~S. Rappaport, Wireless communications: principles and practice, Cambridge
  University Press, 2024.

\bibitem{CSI_overhead}
A.~Zappone, M.~Di~Renzo, F.~Shams, X.~Qian, M.~Debbah, Overhead-aware design of
  reconfigurable intelligent surfaces in smart radio environments, IEEE
  Transactions on Wireless Communications 20~(1) (2020) 126--141.

\bibitem{CSI_2021_large}
N.~K. Kundu, M.~R. Mckay, Large intelligent surfaces with channel estimation
  overhead: Achievable rate and optimal configuration, IEEE Wireless
  Communications Letters 10~(5) (2021) 986--990.

\bibitem{tong2022two_stage}
J.~Tong, H.~Zhang, L.~Fu, A.~Leshem, Z.~Han, Two-stage resource allocation in
  reconfigurable intelligent surface assisted hybrid networks via multi-player
  bandits, IEEE Transactions on Communications 70~(5) (2022) 3526--3541.

\bibitem{RIS_UAV_ergodic_throughput_max}
S.~Lin, Y.~Zou, D.~W.~K. Ng, Ergodic throughput maximization for
  ris-equipped-uav-enabled wireless powered communications with outdated csi,
  IEEE Transactions on Communications 72~(6) (2024) 3634--3650.
\newblock \href {https://doi.org/10.1109/TCOMM.2024.3358563}
  {\path{doi:10.1109/TCOMM.2024.3358563}}.

\bibitem{linear_RIS_CE}
T.~L. Jensen, E.~De~Carvalho, An optimal channel estimation scheme for
  intelligent reflecting surfaces based on a minimum variance unbiased
  estimator, in: ICASSP 2020 - 2020 IEEE International Conference on Acoustics,
  Speech and Signal Processing (ICASSP), 2020, pp. 5000--5004.
\newblock \href {https://doi.org/10.1109/ICASSP40776.2020.9053695}
  {\path{doi:10.1109/ICASSP40776.2020.9053695}}.

\bibitem{CVX}
M.~Grant, S.~Boyd, Cvx: Matlab software for disciplined convex programming,
  version 2.1, \url{http://cvxr.com/cvx} (March 2014).

\bibitem{CE_Multi_UE_RIS}
Y.~Wei, M.-M. Zhao, M.-J. Zhao, Y.~Cai, Channel estimation for irs-aided
  multiuser communications with reduced error propagation, IEEE Transactions on
  Wireless Communications 21~(4) (2022) 2725--2741.
\newblock \href {https://doi.org/10.1109/TWC.2021.3115161}
  {\path{doi:10.1109/TWC.2021.3115161}}.

\bibitem{CE_Multi_UE_RIS_framework}
Z.~Wang, L.~Liu, S.~Cui, Channel estimation for intelligent reflecting surface
  assisted multiuser communications: Framework, algorithms, and analysis, IEEE
  Transactions on Wireless Communications 19~(10) (2020) 6607--6620.
\newblock \href {https://doi.org/10.1109/TWC.2020.3004330}
  {\path{doi:10.1109/TWC.2020.3004330}}.

\bibitem{RIS-performance}
Q.~Tao, J.~Wang, C.~Zhong, Performance analysis of intelligent reflecting
  surface aided communication systems, IEEE Communications Letters 24~(11)
  (2020) 2464--2468.
\newblock \href {https://doi.org/10.1109/LCOMM.2020.3011843}
  {\path{doi:10.1109/LCOMM.2020.3011843}}.

\end{thebibliography}

\end{document}